# Improved Rheometry of Yield Stress Fluids Using Bespoke Fractal 3D Printed Vanes

Crystal E. Owens*, A. John Hart, Gareth H. McKinley

*Department of Mechanical Engineering, Massachusetts Institute of Technology, Cambridge, MA 02139, United States*

*Corresponding Author: rheo3D@mit.edu

**Keywords:** Rheology, yield-stress fluid, vane geometry, measurement tools,

## Abstract

To enable robust rheological measurements of the properties of yield stress fluids, we introduce a class of modified vane fixtures with fractal-like cross-sectional structures. A greater number of outer contact edges leads to increased kinematic homogeneity at the point of yielding and beyond. The vanes are 3D printed using a desktop stereolithography machine, making them inexpensive (disposable), chemically-compatible with a wide range of solvents, and readily adaptable as a base for further design innovations. To complete the tooling set, we introduce a textured 3D printed cup, which attaches to a standard rheometer base. We discuss general design criteria for 3D printed rheometer vanes, including consideration of sample volume displaced by the vanes, stress homogeneity, and secondary flows that constrain the parameter space of potential designs. We also develop a conversion from machine torque to material shear stress for vanes with an arbitrary number of arms. We compare a family of vane designs by measuring the viscosity of Newtonian calibration oils with error <5% relative to reference measurements made with a cone-and-plate geometry. We measure the flow curve of a simple Carbopol yield stress fluid, and show that a 24-arm 3D printed fractal vane agrees within 1% of reference measurements made with a roughened cone-and-plate geometry. Last, we demonstrate use of the 24-arm fractal vane to probe the thixo-elasto-visco-plastic (TEVP) response of a Carbopol-based hair gel, a jammed emulsion (mayonnaise), and a strongly alkaline carbon black-based battery slurry.

## I. INTRODUCTION

A yield stress fluid is a material that has a critical stress above which it flows like a viscoplastic liquid, and below which it deforms as a viscoelastic solid. Common yield stress fluids include emulsions, foams, particulate suspensions, and granular materials, in which particles, bubbles, emulsions, or other microparticle constituents interact via weak physico-chemical forces and geometric packing/jamming constraints [1–4]. As the imposed stress acting on these soft solids increases, complex time-dependent rheological signatures arise from underlying microstructural processes such as shear-induced break-down and restructuring [3,5]. In addition, other effects can arise such as time-dependent aging and onset of non-homogeneous flow, resulting in common rheological signatures, including a strong influence of the history of deformation, hysteresis, thixotropy, shear-banding, and of slip of the material on the surface of the tool used for rheological measurements [6,7]. As a result, sensitivity to loading conditions, ensuring kinematic homogeneity, and unambiguous control of history of deformation each pose particular challenges for rheological measurements of yield stress fluids [5,8].

Despite this complexity, yield stress materials find enormous application in a wide variety of commercial products, consumer goods, and construction materials due to the desirable mechanical properties imbued by the presence of a critical yield stress. Yield stresses in industrial processes influence the strength of concrete, the most utilized artificial material in the world [9], and the processing, quality, and final texture of a vast range of foodstuffs, skincare and haircare products [10–15]. Further, thixotropic materials with high values of the yield stress are particularly suited to high-resolution direct-write printing. Examples include printing of structures using foams, elastomers, concrete, cell-laden gels, and conductive inks [16–21]. To assist in understanding and optimizing yield stress materials for applications, rapid, reliable, and accurate measurements of the material behavior are required. Simultaneously, a key research aim of the field over the past 20 years has been to develop better descriptive and predictive constitutive models that capture the complex rheological behavior of these fluids including description of the full thixo-elastoviscoplastic (TEVP) response (see, for example, [22–24]).

As a result of the complexities encountered in measuring the rheology of TEVP yield stress fluids, the vane has become the rheometric tool geometry of choice, as it prevents the slip of material and minimizes sample damage/alteration during the sample loading process [6]. The vane geometry initially was developed in the 1980's by civil engineers as a tool to quantify the yield stress of soils and thick clays. Seminal work by Nguyen and Boger adapted the vane for muds and slurries and derived a simple quantitative relationship between the torque imposed on the rotating vane and the resulting shear stress acting on the sample [10,25]. The vane subsequently has become a standard tool for measuring the yield stress of delicate materials and structured fluids [6].

The vane geometry most typically consists of four to six straight blades of equal length fanning out from a center point in a cruciform or hexagonal arrangement [6]. The vane is submerged inside a cup of the sample material and then rotated about its central axis; the rheometer records the torque and rotation angle. This rotation deforms an approximately cylindrical plug of material, generating an ideally axisymmetric stress field, while also restricting sample slip, which is a key issue for cylindrical Couette rotors [6]. The vane, despite its more complex geometry, has been used to impose a range of standard rheometric test protocols, including measurement of steady state flow curves, start-up of steady shear, creep/recoil, small amplitude oscillatory shear (SAOS), and large amplitude oscillatory shear (LAOS) [12,21,26]. Systematic comparisons of rheological measurements made using vanes to those using other standard rheometric tools generally have found good numerical agreement between values measured with vanes and with other geometries. Meanwhile, any differences in measured values typically occur due to wall slip and differing sample history, affecting primarily large strain measurements and general level of repeatability in data measured with vanes compared to other geometries. In particular, we note the following specific examples: for bentonite, direct comparisons have revealed that cone-and-plate tests systematically underestimate yield stress as compared with vanes, due to thixotropy associated with sample loading [27]; for foams, SAOS measurements with vanes agree well with measurements made with parallel plate fixtures, while vanes induce less bubble coalescence [28]; for soft cheeses, SAOS measurements with a parallel plate geometry agree well with vanes, though only vanes were capable of imposing reproducible larger-amplitude strain deformations due to sample slip occurring against the parallel plates [29]. In direct comparisons of inter- and intra-laboratory tests, yield stress measurements made with vanes in start-up of steady shear tests have been found to be more reproducible than measurements made by other tools including slump tests on inclined planes, or creep experiments or stress ramp tests with cones and textured concentric cylinders [30].

The shearing profile around the vane is axisymmetric only for certain materials and under specific flow conditions. Secondary flows arise when the viscosity is too low, resulting in recirculation between neighboring pairs of arms [31,32]; these are exacerbated when the power-law index of a shear thinning fluid is >0.5 (more Newtonian) [33]; and when the vane has too few arms to hold a given material securely (typically needing greater than three arms) [34]. For viscoelastic materials, this recirculation has been shown to cause a significant artificial increase in the apparent viscosity reported by the instrument [35]. Detailed theoretical and computational analysis of the region near a single knife-edge of a vane reveal that the stress field is singular at the tip, and consequently the stress field around a multi-arm vane tool shows spatially periodic variations for any material with Newtonian, yield stress, or linear elastic behavior [31,36,37]. Consequently, even when the streamlines in a sheared fluid sample are circular, instantaneous structural parameters characterizing the local properties of thixotropic fluids can be strongly influenced by the location of the blades, becoming non-axisymmetric with the vane rotation [37].

Despite these difficulties, vanes are particularly useful for ensuring repeatability of measurements and for characterizing structurally-sensitive materials. This is because vanes displace far less material and impose a much weaker deformation history during initial sample loading, which is particularly important for thixotropic samples [38]. With a cone-and-plate, parallel-plate, or concentric-cylinder tool configuration, the sample must be compressed and sheared to fill the thin gap between the two fixture surfaces. Compared to a bob, a vane can more easily be inserted into a cup that has been previously filled with a structured material, and the vane typically displaces less than 20% of the sample volume compared to a bob of equivalent radius. This results in more repeatable measurements and control of the material's initial shear history. Furthermore, use of vanes allows samples to be prepared and aged in containers for long waiting times before testing on the rheometer [27].

Alternatives to the vane geometry that have been proposed for yield stress fluids include paired helical blades for preventing sedimentation while measuring dense samples such as concrete with large aggregates, torsional mixers, and planetary rotating systems [39], along with other styles of test including penetration and slump tests [5,40]. Due to their easy insertion into fluids, four-armed vanes also find wide-spread use in field tests for industrial measurements, with designs for "bucket rheometers" for concrete and industrial slurries powered by a hand drill [41] and a similar extended rod for *in situ* or "*syn-eruptive*" measurements of lava flows oozing from active vents, where the magnitude of the yield stress is a strong indicator of probability of eruption [42,43].

In the interest of allowing easier design and fabrication of rheometric tooling, Bikos and Mason recently introduced 3D-printed (3DP) cones and annular rings for rheometers as a cost-effective approach to create bespoke parts [44]. Other researchers have used 3DP to make custom drag-reducing surfaces for viscous skin friction tests utilizing the rheometer motor and torque sensor [45]. In other fields, 3D printing, a subset of a class of processes broadly known as additive manufacturing, has become widespread for production of complex geometries from a vast library of possible base materials [46]. In the present case, 3DP is well-suited to create functional vane-like geometries, compared to other manufacturing methods, due to the need to produce fine (<1mm) features with very high aspect ratios that retain dimensional accuracy over O(cm) length scales [47].

In this paper, we present the design, fabrication, and use of a 3D printed fractal-like vane geometry. The branching, tree-like fractal structure was optimized to give a large surface area and large number of contact edges with the test fluid, while the internal structure remains sparse in terms of displaced volume relative to a bob, in order to limit pre-shearing of a structurally-sensitive material during sample loading. In Section II, we discuss the relevant stress and strain fields for a generic vane geometry, combining and adjusting published expressions to propose a composite formula for converting torque to stress, incorporating variations in the number of arms, the vane geometry, and the influence of end effects from a vane with finite length. We also discuss the impact of design variations on the projected area displaced by the vane geometry, and on the stress profile around the vane geometry. In Section III, we discuss the design and manufacturing of vanes by stereolithographic 3D printing, using a methacrylate-based photopolymer to create vanes with fine (200 µm) feature resolution and with chemical compatibility with a broad range of solvents and sample materials. We also discuss a printable coupling to a common rheometer interface, and propose a design for textured 3D printed cups that assists in preventing sample slip at the outer walls, while also being detachable from the rheometer base for easy cleaning. In Section IV, we quantify the accuracy of the new designs by comparing vane measurements of Newtonian calibration oils and a simple (*i.e.,* non-thixotropic) yield stress material (a Carbopol microgel) with reference measurements obtained in a roughened cone-and-plate fixture, and compare these results to our proposed torque scaling factors. In Section V, we use these vanes to perform start-up of steady shear flow measurements of the TEVP response of a Carbopol-based hair gel, a jammed emulsion (mayonnaise), and an alkaline carbon black-based battery slurry with pH 12. Following our conclusions in Section VI, we provide four appendices discussing in more detail (A) the stress field around the new vane geometries; (B) the definitions and selection of particular fractal structures; (C) the displaced area filled by vanes of different designs; and (D) how to ensure a vane meets acceptable standards for accurate and reproducible rheological measurements, with the goal of facilitating adoption of our design and fabrication method in other laboratories.

## II.    THEORY

### A.  Rheology of yield stress fluids

The generic term "yield stress fluid" typically applies to a fluid that exhibits a characteristic stress below which it may deform viscoelastically (*i.e.,* as it creeps) but does not flow, and above which it flows steadily like a (typically shear-thinning) liquid. For further details, see the extensive reviews provided by [3,6,48,49]. In addition to the key role of a critical material stress, time-dependent degradation and rebuilding of the underlying material structure may occur, leading to complex rheological responses that depend both on time and on sample history.

The simplest constitutive model appropriate for describing the steady flow curve of yield stress fluids is the Hershel-Bulkley model, [50]

$$\begin{aligned} \sigma(\dot{\gamma}) &= \sigma_y + k\dot{\gamma}^n \quad & \sigma \geq \sigma_y \\ \dot{\gamma} &= 0 \quad & \sigma < \sigma_y \end{aligned} \tag{1}$$

where $\sigma$ is the shear stress, $\sigma_y$ is the yield stress in shear, $k$ is the consistency index, $\dot{\gamma}$ is the shear rate, and $n$ is the power law index. When $n = 1$, this corresponds to a Bingham fluid, with $k \rightarrow \mu$ being the plastic viscosity. While many more complex models exist which can account for viscoelastic responses below yield as well as time and rate-dependent thixotropic responses (*ex.*, the soft glassy rheology model [24], the isotropic kinematic hardening model [23,26], and models by Saramito [51,52] and Coussot [53]), the simple Hershel-Bulkley viscoplastic model is sufficient for comparison with the results of steady state measurements presented here.

### B.  Torque-shear stress and rotation rate-shear rate relations for the vane geometry

The typical vane (Fig 1a) has four arms connected in a cruciform design; it is inserted centrally into a cup of material and rotated about its axis while the torque is measured as a function of rotation angle and rate. Due to the presence of a yield stress in the material being measured, the sample in the cup deforms as a sheared cylinder guided by the vanes, preventing slip and generating kinematics that closely approximate those of a concentric cylinder system. The essential question that arises is how to relate the global values measured by the rheometer (*i.e.*, the torque $M$, yield torque $M_y$, and rotation rate of the vane $\Omega$), to constitutive variables $\sigma$ and $\dot{\gamma}$ introduced in equation (1) so that the three model parameters $\sigma_y$, $k$, and $n$, can be calculated.

In a vane measurement, an *N*-arm vane with radius $R_v$ rotates inside a cup of radius $R_c$ at a rotation rate $\Omega$. There are two regimes of material response for yield stress fluids. When the applied torque is below the yield torque, $M < M_y$, or correspondingly $\sigma < \sigma_y$, the sample is entirely plastically unyielded, although a transient viscoelastic creep may occur throughout the entire sample when the torque is first applied. Due to the radial dependence of the true strain in the material, it is useful to define an "apparent strain" which can be calculated as the observable angle rotated by the vane through a time $t$

$$\theta_{app} \approx \int_0^t \frac{R_v \Omega(t')dt'}{R_v} \approx \theta(t) \quad \text{(for } \sigma_w < \sigma_y) \tag{2}$$

and a rotation rate $\Omega(t)$, where the wall shear stress $\sigma_w = \sigma(R_v)$.

The conversion of torque to stress depends purely on the test geometry. In their original analysis, Nguyen and Boger used a cylindrical model for the yield surface and calculated the stress on the cylindrical yield surface enclosing the vane including the cylindrical wall and both end caps to the total torque:

$$M = 2\pi R_v^2 \sigma_w \int_0^L dz + 4\pi \int_0^{R_v} \sigma(r)r^2 dr, \tag{3}$$

where *L* is the finite length of the vane. The additional contribution from the thin rod holding the vane is neglected here, as our vane has $(R_{rod} / R_v)^2 \approx 0.1$ and $(L_{rod} / L_v) \approx 0.17$. Usually, one assumes $\sigma(r) = \sigma_y$ everywhere on the two end cap surfaces to integrate equation (3), as in [10], to obtain

$$M = \pi R_v^3 \left( \frac{L}{R_v} + \frac{2}{3} \right) \sigma_w \tag{4}$$

This relationship enables an interconversion between the observable (torque) and the desired rheometric variable, *i.e.,* the wall shear stress at the surface of the vane.

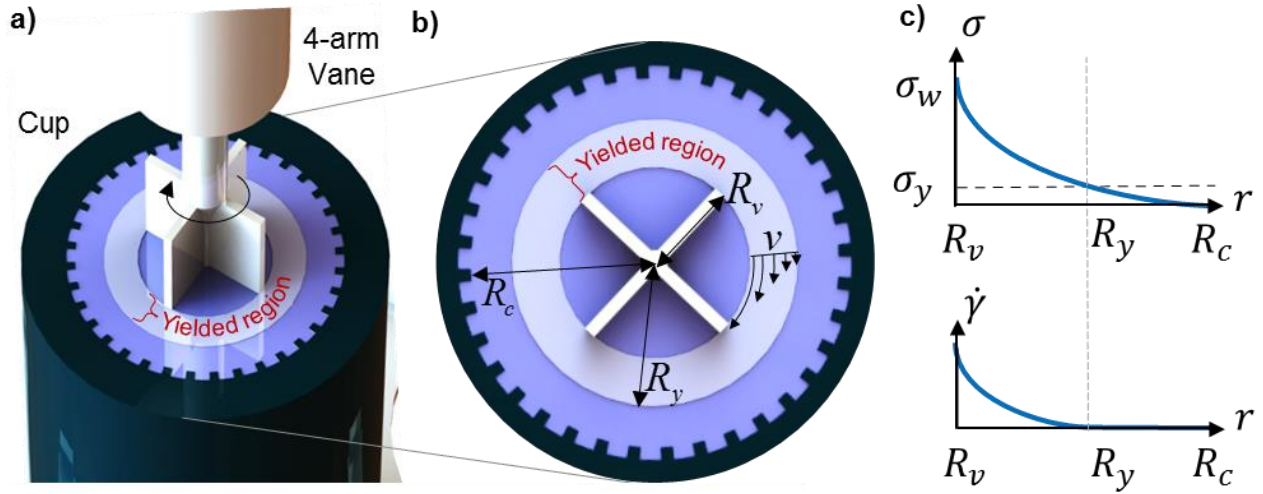

**FIG. 1**. (a) Schematic of a 4-arm vane ($N = 4$) inserted halfway (for illustrative purposes only) into a cup of yield stress material (blue) for measurement. As the vane rotates, the shearing of material exerts a torque on the vane that is measured by the rheometer. (b) The local stress decays radially from the cylinder of material cut by the vane, out to a radius $R_v$ where $\sigma(r) = \sigma_y$. Inside the vane boundary $r < R_v$, material ideally moves as a solid plug guided by the vane arms, and the radial velocity $v(r)$ decays from $v = \Omega R_v$ at the edge of the vane to $v = 0$ at $r = R_y$. (c) The shear stress field in the sheared region of fluid decays from a wall shear stress $\sigma_w$ at $r = R_v$ to the yield stress $\sigma_y$ at a critical radius $R_y$, beyond which material is unyielded. The shear rate $\dot{\gamma}(r)$ of material elements also decays to zero at $R_y$.

When $M > M_y$, or $\sigma_w > \sigma_y$, the sample begins to yield and eventually rotates at a constant rotation rate $\Omega$. If the flow is assumed to be purely axisymmetric (due to conservation of angular momentum), the solution of the Cauchy momentum equation shows that the shear stress field in the sample decays radially as $\sigma \sim 1/r^2$ until the shear stress $\sigma(r) = \sigma_y$ at a critical radius $R_y = R_v \sqrt{\sigma_w / \sigma_y}$, beyond which $\dot{\gamma} = 0$ by definition (Fig 1b-c). The surface at $r = R_y$ is called the yield surface. In the *partially yielded case*, the outer edge of the plastically flowing material does not reach the outer wall of the cup ($R_y < R_c$). In the *fully yielded case*, the fluid is plastically yielded throughout the cup, and $R_y \rightarrow R_c$ and this necessitates the use of a different analysis [25]. In the present study, the chosen gap is large enough (or, equivalently, the applied torque is maintained at small enough values) that the fluid is only partially yielded for all measurements, which means that $\sigma_{w,\max} / \sigma_y < (R_c / R_v)^2$. We note that the yield surface reaches the outer wall in our system when the shear stress computed from equation (4) reaches a threshold value

$$\sigma_{w,\max} \cong \sigma_y \left( \frac{R_c}{R_v} \right)^2 \qquad (5)$$

and we demarcate this upper bound as a dashed line on the relevant figures in this article. For all results presented here, $(R_c / R_v)^2 \geq 4$ so that we can often measure the entire flow curve accessible with the rheometer using the equations presented above.

A wide gap system is defined as one in which the gap between the vane and the cup, $(R_c - R_v)$ is on the same scale as the vane radius, $(R_c - R_v) / R_v \sim O(1)$. The shear rate $\dot{\gamma}(r)$ in the material is related to the rotation rate of the vane by the identity $\dot{\gamma}(r) = \frac{1}{r} \frac{d}{dr}(r v_\theta)$. As described by [25], this expression can be rearranged and integrated to obtain:

$$\Omega = \int_{R_v}^{\infty} \dot{\gamma} \frac{dr}{r} = -\frac{1}{2} \int_{R_c}^{R_v} \dot{\gamma}(R_v) \frac{d\sigma}{\sigma} . \qquad (6)$$

The relationship between rotation rate and stress depends on the specific (*a priori* unknown) constitutive response of the fluid sample in the gap, and not on geometry alone. However, when $R_v < R_y < R_c$, this implicit equation can be solved without presuming any specific fluid model (besides the presence of a critical yield stress) by realizing that $\dot{\gamma}(R_y) \to 0$. Differentiating equation (6) by $\sigma$, evaluating it at $r = R_v$ where $\sigma = \sigma_{wall}$, and substituting the relationship between wall shear stress and torque derived in equation (4) then gives the shear rate at the vane surface to be

$$\dot{\gamma}(R_v) = 2\sigma \left( \frac{d\Omega}{d\sigma} \right) \Big|_{\sigma = \sigma_w} = \frac{2\Omega}{\left( d \log M / d \log \Omega \right)} \qquad (7)$$

This expression depends on the actual stress-shear rate (or torque-rotation speed) relationship of the material being tested, and so only a prescribed rotation rate and not a prescribed shear rate may be imposed *a priori* for experiments on an uncharacterized material. Other expressions for calculating $\dot{\gamma}(r)$ from the imposed rotation rate $\Omega$ when the entire sample is yielded (*i.e.*, when $R_y \geq R_c$), or for materials without a yield stress are compared in [25,32,54]. For the specific case of measuring a viscous Newtonian fluid, the wall shear rate, $\dot{\gamma}_w$, at the perimeter of the vane can be evaluated as [25]

$$\dot{\gamma}_w = \frac{1}{r}\frac{d}{dr}\left(r v_\theta\right)\bigg|_{r=R_v} = \frac{2\Omega}{1-\left(R_v / R_c\right)^2} \qquad (8)$$

## C. Generalized torque-to-stress conversion relations for vane geometries

While the vane geometry is well suited for measuring the instantaneous torque $M_y$ (and the corresponding yield stress $\sigma_y$ at the instant of yielding), the streamlines become noncircular when measuring strongly shear-thinning materials at higher shear rates, or for measuring viscous Newtonian fluids, due to secondary flows (*i.e.*, slow recirculation of fluid eddies) between the vane arms [33]. Essentially, the fluid-filled space between each pair of vane arms functions as a lid-driven cavity with a slow steady recirculating flow that contributes additional dissipation to the total measured torque. A design that occludes more internal space may be more suitable for these measurements, as was recognized even in early work by Keentok et al: "*A more theoretically acceptable vane would be designed in such a way as to eliminate the possibility of secondary flows between the blades.*" [34]. Yet, this requires an intricate vane geometry (*e.g.,* containing many thin blades), which is expensive to mold or machine, making their manufacture cost-prohibitive at that time.

In addition, the stress field in the sheared sample is non-uniform and localized around the perimeter at the edges of each arm of the vane. In fact, analysis by Atkinson and Sherwood for the stress field (in a Newtonian fluid or a linear elastic solid) near a knife-edge singularity in torsional deformation shows that at each arm tip, there is a stress singularity that scales inversely proportionally to the number of vane arms [31]. By increasing the number of tips in contact with outer fluid, the stress field becomes progressively more homogeneous along circles of constant radius from the vane center. To illustrate this, we evaluate the azimuthal shear stress $\sigma_{r\theta}$ field for a Newtonian fluid along the line at $r = 1.05R_v$ as shown in Fig 2a, b. The azimuthal shear stress profile ($\sigma_{r\theta}$) is plotted in Fig 2b along the angular spacing between two adjacent vane arms, and in Fig 2c the same stress is plotted for each vane. Additional details of this solution structure are provided in Appendix A.

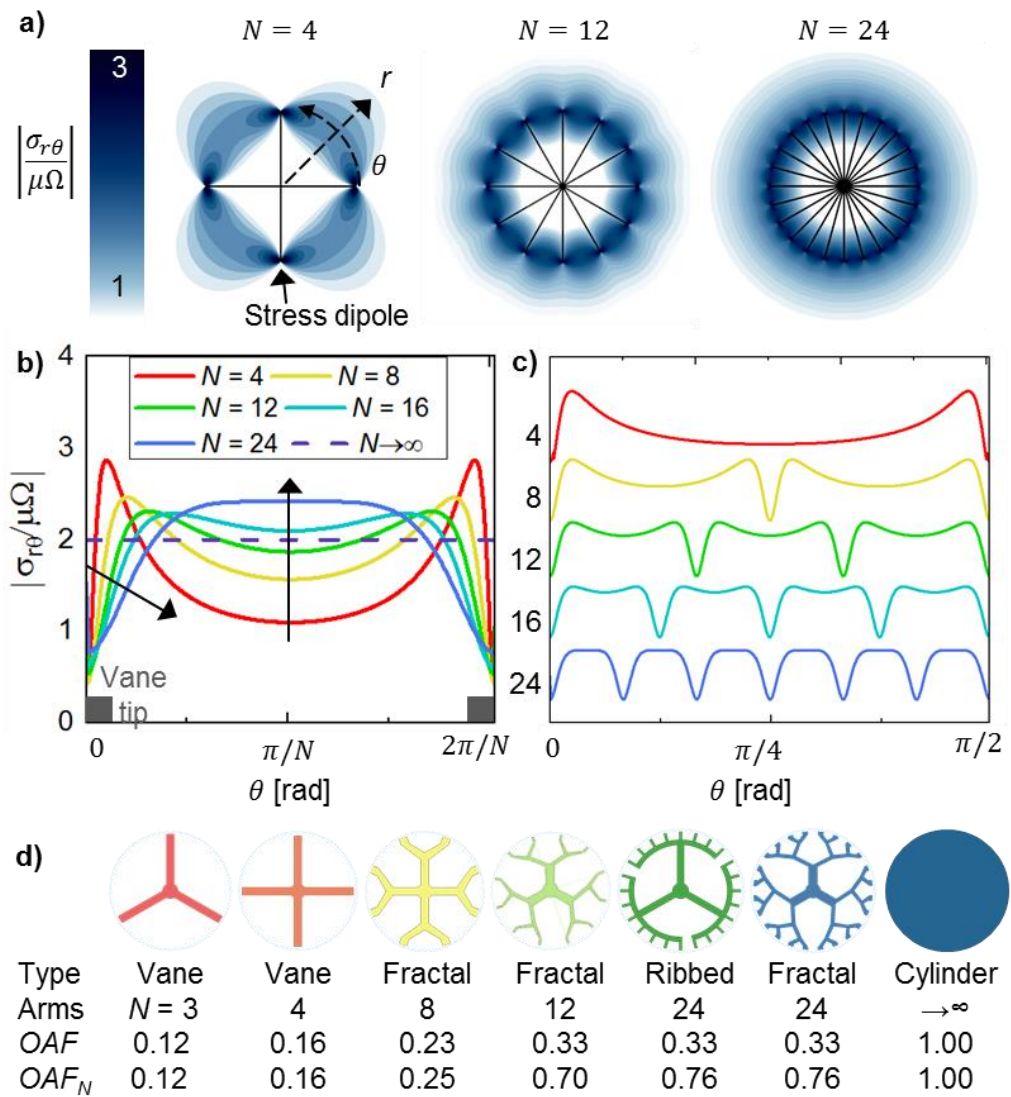

**FIG. 2**. (a) Analytical contour plot of the shear stress field for a Newtonian fluid calculated using equations (A1-A3) around a 4-armed vane in an unbounded domain, normalized by $\mu\Omega$ as a representative scale for the viscous stress. Strong dipolar stress concentrations are present at each vane tip. Progressively smoother profiles form around 12 and 24-armed vanes. (b) The azimuthal shear stress profile $\sigma_{r\theta}(\theta)$ as a function of angle $\theta/N$ between two neighboring vane tips for a family of vanes with increasing number of arms, calculated at a constant radius of $r = 1.05R_v$. The first vane arm is located at $\theta = 0$. The stress initially increases sharply due to the stress dipole, and then decreases between arms. (c) Profiles for $0 < \theta < \pi/2$ for the same vanes in (b), shifted vertically for clarity. The y-axis labels indicate the number of arms. (d) A series of vane designs considered in the present article, along with the number of vane arms $N$, the occluded area fraction (OAF) displaced by the solid vane compared to an equivalent circle, and a comparison of the area occluded by the vane compared to the corresponding area of a vane with $N$ straight arms (denoted $OAF_N$).

As the vane rotates, it creates a dipolar stress field emanating from each vane tip. As a result, the minimum values of shear stress are attained on lines emanating radially outward from the vane tip ($r \approx R_v, \theta = 0$ in the coordinate system shown in Fig 2(a)). Moving away from the arm tip along the direction of increasing angle $\theta$, the stress initially increases sharply, and then decreases to a local minimum in the gap between arms. As the design is changed to incorporate a greater number of vane arms, the background shear stress between the arms approaches an increasingly homogeneous plateau value. Evaluating equations (A1-A3) in Appendix A for a Newtonian fluid, with $r = R_v$ and $0 \leq \theta \leq \pi/2$ we find that the stress drops by 60% for a 4-armed vane, by 18% for a 12-armed vane, and has no drop between arms but increases 2% for a 24-armed vane.

Although the analysis leading to equations (2)-(4) discussed above considers the torsional stress field in the test material to be axisymmetric, in reality the shearing stress varies in $r$ and $\theta$ and is spatially localized at the outermost contact edges between the vane and the fluid as indicated in Fig 2(a). In the 1990s, Sherwood *et al.* [36] simulated the influence of the number of arms, $N$, on a vane tool as well as end effects arising from 3D flow and showed that the measured torque, $M$, and the wall shear stress are related more accurately by the following expression by:

$$M = 2\pi R_v^2 L \sigma_w \left[ \left( 1 - \frac{1}{N} \right) + \frac{R_v}{4L} \left( 2.75 - \frac{3}{\sqrt{N}} \right) \right]. \tag{9}$$

It is evident from this relationship that increasing the number of vane arms brings the stress closer to the homogeneous wall shear stress expected for a cylindrical bob ($N \rightarrow \infty$), and increasing the vane aspect ratio ($L/R_v$) reduces the contribution of the end caps of the vane to the total torque. Later, Atkinson et al. derived a similar relationship between the total torque and the wall shear stress for a vane with an arbitrary number of arms by using a 2D analysis for the stress field in the vicinity of a knife edge (thin plate) undergoing torsional deformation, which is valid when $L/R_v \rightarrow \infty$. Detailed enumeration of the integrals resulted in an expression for the torque per length on a 2D representation of an $N$-vane cross-section, which for a vane of length $L$ gives a total torque [31]:

$$M = 2\pi R_v^2 L \sigma_w \left( 1 - \frac{1.113}{N} \right). \tag{10}$$

We have combined and adjusted this 2D analytical solution with the 3D simulation result, guided by experimental data obtained with silicone oils and a simple yield stress fluid. We use these to develop two practical stress-torque conversion factors (denoted $S_\sigma$) based on geometric features of the vane geometry that we can then compare with our experimental data. From equation (4), we can see that if we assume a **cylindrical stress profile expression**, then the appropriate torque conversion factor ($S_\sigma^O$) is given by:

$$S_\sigma^O \triangleq \frac{M}{\sigma_w} = 2\pi R_v{}^2 L\left(1 + \frac{2R_v}{3L}\right).$$

(11)

We expect this expression to perform well when the deformation in a sample is nearly ideally cylindrical (for example at stresses below and close to the yield stress, and for vanes with a large number of arms). By combining the scaling of equation (10) and the numerical calculations underpinning equation (9), we also construct the following **N-dependent expression** for the stress conversion factor:

$$S_\sigma^N \triangleq \frac{M}{\sigma_w} = 2\pi R_v{}^2 L\left[\left(1 - \frac{1.113}{N}\right) + \frac{R_v}{4L}\left(2.75 - \frac{3}{\sqrt{N}}\right)\right].$$

(12)

We expect this expression to be more accurate for structurally-sensitive materials with rheology that depends on the local kinematics near each vane. By considering the typical reproducibility of experimental rheometric data with yield stress fluids and enumerating equations (11) and (12) for a range of N (with $L \gg R_v$), we find N = 24 to be the lowest even-numbered vane geometry in which the difference between the two expressions varies by less than 5%, and this is the largest value of N used in the present study.

However, simply adding more arms also displaces and disturbs more material when the vane is inserted into a sample, which has been shown to lead to underestimations of the yield stress [55], and may make measurements on thixotropic fluids that exhibit a strong memory of their initial deformation history during loading less repeatable. To quantify this effect and explore it systematically, we define the occluded area fraction (OAF) as the cross-sectional area of fluid displaced by an N-arm vane normalized by the area of a circle with the same outer radius:

$$OAF \triangleq A_{vane} / \pi R_v^2.$$

(13)

Computing numerical values of this expression obviously depends on $N$ as well as the thickness $t$ of each vane. For example, a 4-arm vane with a thickness of $0.13R_v$ (typical of commercial vanes) has an *OAF* of $4 \times 0.13 / \pi = 0.16 = 16\%$ .

Our design objective, then, is to reduce recirculation of fluid between the vane arms and increase the shear stress homogeneity in the sample close to the vanes without substantially increasing the occluded area fraction of the vane. For this purpose, we propose a fractal design. Certain classes of fractals have well-defined recursive branching structures that lead to particularly high surface area-to-volume ratio [56]. In particular, we explore finite Bethe lattice-like fractals with $N = 12$ and 24 arms, and compare these new designs to classical designs for vanes with $N$ radial arms, and a ribbed cylinder textured with 24 ribs. Cross-sectional profiles of each of these designs are shown in Fig 2(d) along with numerical values of OAF for each design, and corresponding values for vanes with $N$ straight radial arms (denoted OAF$_N$ for clarity). These designs are discussed further in Appendix B.

The area minimization problem was approached through analysis of the area occluded by each profile. The different design families we compare were guided by results from a pre-programmed Steiner tree algorithm [57], which generates the fully optimal, global minimum length path spanning a set of input points. In this case, the input design was a set of $N$ points spaced evenly around the perimeter of a circle with radius $R_v$ plus one additional point at the center (corresponding to the spindle location). These computational results revealed the sparsest network connecting the $N+1$ points along with the lower bound of the achievable *OAF* for a given $N$ (and fixed arm thickness, $t$). Further details of this analysis are included in Appendix C.

## III.    DESIGN AND 3D PRINTING OF VANES AND RHEOMETER TOOLS

## A.  Design of vane for stereolithography

Rheometric tools are typically manufactured via machining of aluminum or of stainless steel. The introduction of a new fractal-based vane geometry requires a manufacturing method that can achieve complex features without excessive cost, and that is capable of creating thin and closely-spaced features such as the profiles shown in Fig 2(d). 3D printing is exceptionally well suited for this task, as its layer-by-layer nature allows complex geometries to be created in three dimensions with, ideally, minimal post-processing or shaping. In particular, designs such as these extruded vanes that are geometrically complex along a single axis and without overhanging features can be printed with high quality using stereolithography [47]. Manufacturing thin, high-aspect ratio cavities by other methods such as machining, injection molding, or casting would be more challenging [58].

We selected the overall dimensions and rheometer coupling of our fractal vanes based on the commercial vane ($N = 4$) available for the TA Instruments DHR series of rheometers with a diameter of $2R_v = 15$ mm and a length of $L = 30$ mm. The vane body was printed via stereolithography on a Form2 printer (Formlabs, Inc.) using Clear resin (Formlabs, Inc.), a methacrylate-based translucent photopolymer with Young's Modulus $E = 1.6$ GPa, and ultimate tensile strength UTS = 38 MPa (pre-curing) [59]. To facilitate attachment of the vane directly to the draw rod of the rheometer, we included a threaded coupling as shown in Fig 3. The 45° slope indicated in Fig 3(a) on the mating joint between the spindle holding the vane and the coupling to the rheometer allowed the vane to be printed entirely without a supporting structure, increasing printing speed and quality, as well as obviating the need for support removal. To enable this, the vanes were printed vertically, vane-end first (Figure 3c). We determined that this orientation ensured the best concentricity of the vane geometry with the printed rheometer coupling and best surface texture as it avoided consecutive layers giving a "stair stepping" texture on the surface and negating any imprecision in the x-y and z-stage calibrations (Fig 3d). The feature resolution is found to be 200 μm over several cm (Fig 3e, f), and vanes were printed with the coarsest 100 μm layer height setting of the Form2 printer. Printing took 3-3.5 hours per single build platform (*i.e.,* per tray shown in Fig 3(c)). The print time scales sublinearly with the number of vanes printed in each tray due to the high speed of lateral in-plane motion that can be achieved in stereolithography, compared to the time required to recoat resin and incrementally move the build platform after each layer. For example, we printed one vane in three hours and 12 vanes in seven hours.

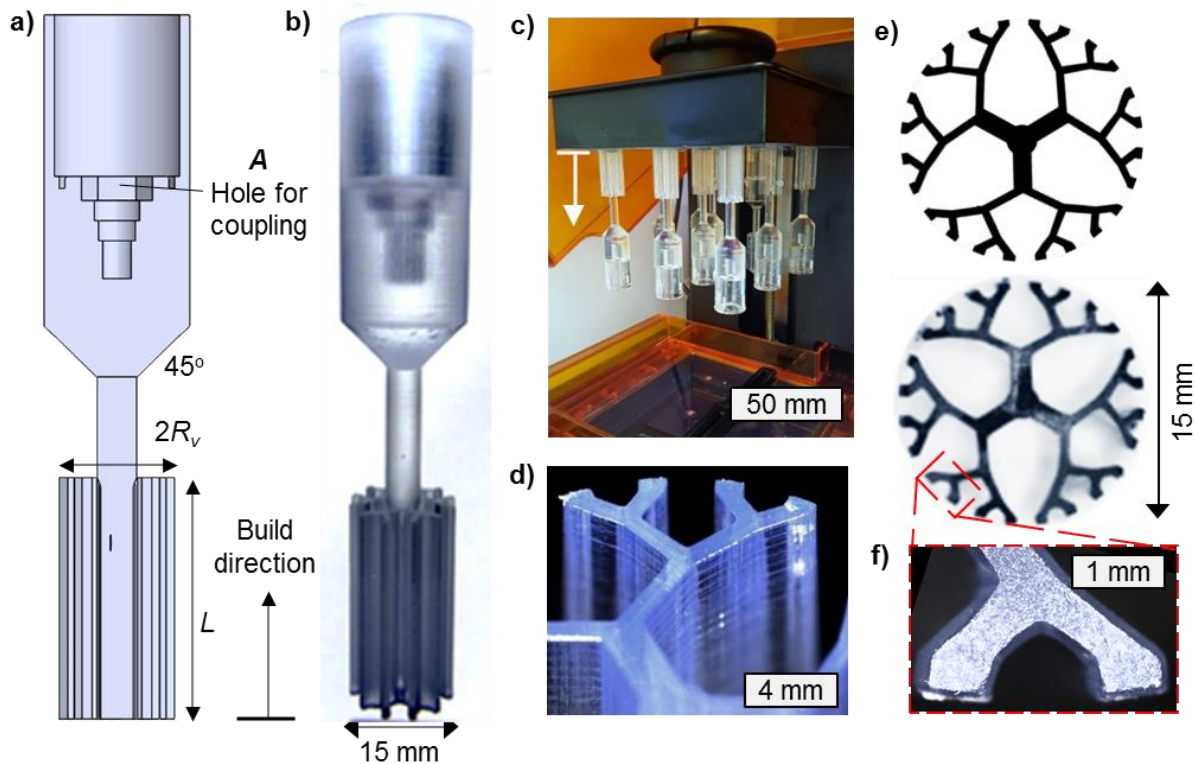

**FIG. 3**. a) A complete vane design for the case of a finite 3-generation Bethe lattice with $N = 24$ contact points at $r = R_v$ and (b) the printed vane mounted on a DHR-3 rheometer spindle. The mounting design (A) allows us to directly connect the spindle coupling to the chosen vane geometry. A 45° transition angle and otherwise straight cylindrical features allows the vane to be printed via stereolithography without supports, building vertically from the vane end first, which is the upper surface in the 3D printer tray shown in (c). An oblique view shows smooth sidewalls despite the presence of visible stacked layers (d). This is because of the chosen printing orientation. (e) The input fractal shape design and a photograph of the fabricated part on end. (f) An optical micrograph of one edge of the printed design shows the ~200μm feature resolution achieved with 3D printing.

Stereolithography, which locally cures photopolymer resin via a light-initiated chemical reaction, allows the use of polymers with wide chemical compatibility [60]. The methacrylate blend material we selected is compatible with solvents such as acetone, mild acids, and also strong bases (including at least one carbon black-based battery electrolyte with pH 12 that corrodes aluminum). The printed vanes have a cost of material of ~$1.90 (including the metal threaded insert). We previously determined the repeatability of stereolithography under similar printing parameters is 30 μm standard deviation in dimensions between prints [61]. Thus, even in the case in which the material to be tested slowly degrades and/or adheres to the vane, 3D printing is still

an attractive solution. Relative to the value of an accurate measurement, the low cost of 3DP vanes may justify their single-time use.

## B. Design of the coupling for attachment to the rheometer

In order to robustly attach to a commercial rheometer, the printed vane geometry was designed to accommodate a threaded insert, rather than by including printed threads integral to the vane. The coupling (A) shown in Fig 3(a) was designed to be 3D-printed, fitted with an M4 threaded insert, and attached directly to a DHR3 or AR-G2 rheometer spindle (Discovery Hybrid Rheometer 3 or Advanced Rheometer-Generation 2; TA instruments, New Castle, DE). These two generations of controlled-stress rheometers have a drag cup motor, radial air bearings, magnetic thrust bearings, and an optical encoder in the head of the rheometer. The stationary base has a temperature sensor and Peltier plate assembly (Fig 4a).[62] A standard geometry slides up onto the spindle and is held axially in place by a long threaded rod (Fig 4b).

The printed vane is designed to fit over the rheometer spindle with slight interference (0.05 +/- 0.02 mm diametrical interference), and a hole to position a threaded nut insert to hold the part in place (Fig 4c-e). The diametrical runout $\Delta R_\theta$, or eccentricity of rotation (a.k.a. "wobble"), was measured using a laser-based line profilometer (LJ-V7080, Keyence) as $0.35 \leq \Delta R_\theta \leq 0.70$ mm total runout at the lowest end for printed vanes, compared with 0.31 mm for a commercial metal vane (TA Instruments, Part 546027.901). Note that the inner diameter of the printed coupling may change slightly when exposed to solvent during the post-print cleaning step, affecting the amount of interference.

For the vane-rheometer coupling, a variety of designs were compared, including five types of threaded inserts and various combinations of dimensional interference and structural compliance; these are shown in Appendix D. The final design using a press-fit nut as a threaded insert (shown in Fig 4e; McMaster press-fit nut, 99437A145), was selected for consistently giving the lowest runout on repeated attachment/detachment of the tool. In addition, the specific nut chosen had a flat top surface, and this surface was pressed flush against the rheometer spindle, and likely contributes to the lower runout achieved. Couplings could easily be designed to fit other rheometers available in the marketplace, and in the supporting information we provide the design of a vane coupling to ARES strain-controlled rheometer with this article (see E.S.I.).

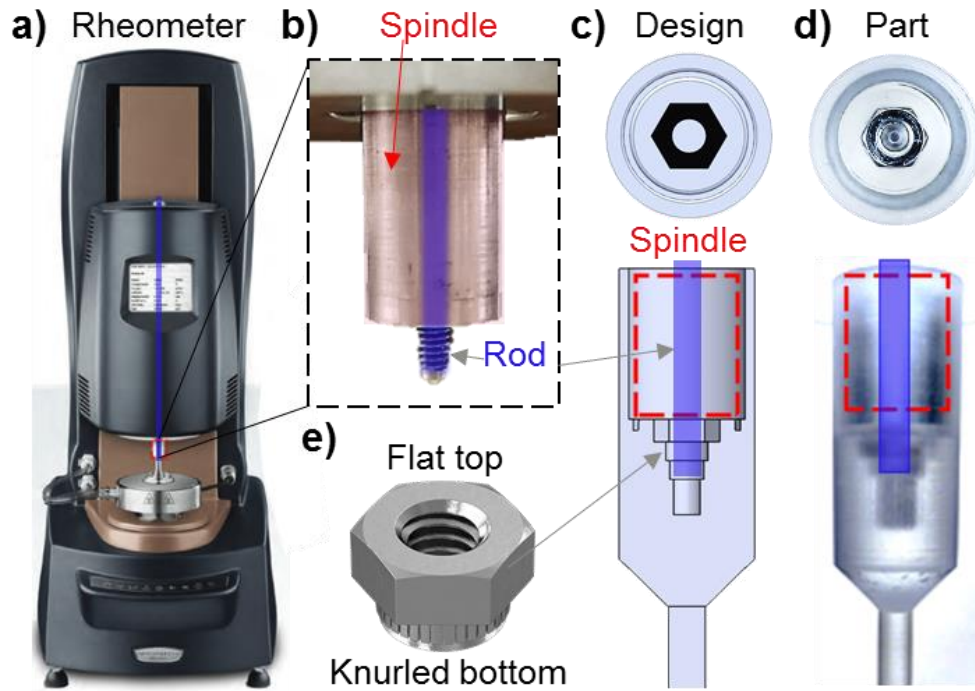

**FIG. 4.** (a) The DHR and AR-G2 family of controlled stress rheometers have a loose threaded "draw rod" extending through the drag cup motor housing and sensing head (outlined in blue), with (b) a spindle (highlighted red) that aligns the geometry to the axis of rotation of the rheometer. (c) An on-end view of the as-designed coupling and (d) a photograph of the 3D-printed part, which has a loose interference fit (50μm overlap) on the spindle for radial positioning, and locks in place axially using the spindle M4 screw with a threaded nut insert. (e) Computer drawing of the press-fit M4 nut that connects the part to the rheometer. The knurled bottom surface holds it in place within a centered hole in the printed vane.

Further, the rotational inertia of a 3D printed acrylic vane ($I_v \simeq 0.8$ μN.m.s$^2$) is much smaller than that of a metal vane ($I_v \simeq 6$ μN.m.s$^2$), although both moments of inertia are much smaller than that of the rheometer head itself ($I_{inst} \simeq 18.5$ μN.m.s$^2$), so this has a negligible effect on most measurements.

## C.  3D-printed cup with textured wall

We also designed a cup to complement the 3D printed vane, and serve as the sample holder. The inner surface of the cup is textured to help prevent slip of the sheared sample at the outer wall, with square steps of 1 mm width by 1 mm depth, spaced apart by 2 mm. These dimensions were chosen to give similar spacing as the arms of the 24-armed fractal vanes and to promote infill of the material into the gaps between ribs (Fig 5a, b). In addition, the cup fastens onto the Peltier plate that is a standard lower fixture on the DHR or AR-G2. The Peltier plate holds it either by an interference fit of six protruding arms or by using the arms as guides and using double-sided tape under the base for a more adjustable hold, as a well-calibrated Peltier plate is orthogonal to the axis of the rotating spindle but is not necessarily concentric with it. The yielded area is set entirely by the location of the vane perimeter and extends radially outwards. Provided the yielded area does not reach the outer wall (*i.e.*, provided the imposed stress is $\sigma_w / \sigma_y < \left( R_c / R_v \right)^2$), the cup shape and position only influences the spatial homogeneity of the linear viscoelastic deformation that occurs pre-yielding, with an effect proportional to the offset of the rotation axis (Appendix D). Thus, for measurements of yield stress fluids with a wide-gap geometry, precise concentricity of the cup is not critical.

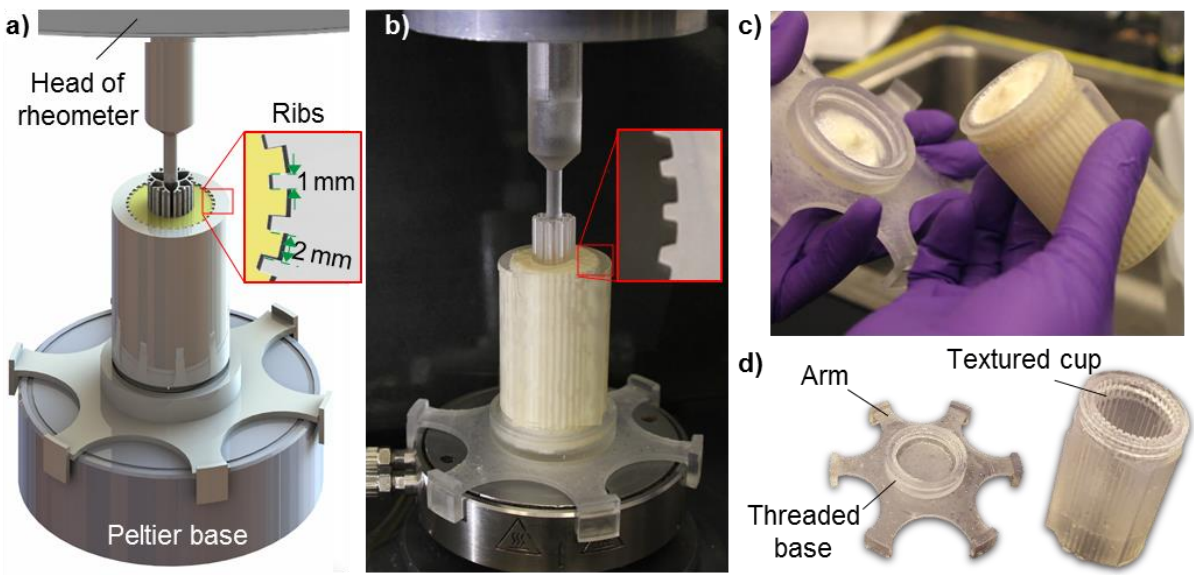

**FIG. 5**. (a) A 3D-printed cup featuring a textured inner wall to prevent slip. Lower arms clip onto the standard Peltier plate base, and the cup walls disassemble from the base via a threaded mating feature for ease of cleaning out thick yield stress materials from within the ribbed surface texture. The inset shows the crenelated ribs around the outer surface, which have 1 mm width and side length, and 2 mm spacing. (b) The assembled cup mounted onto a Peltier plate base, filled with mayonnaise, and (c) the disassembled cup with material still inside, and (d) a cleaned and disassembled cup and base.

The cup was designed for facile use with yield-stress fluids, which often are challenging to fill into the cup without bubbles and to clean out after an experiment, especially given the crenelated walls. As a result, the cup was designed to unscrew from the base, making it straightforward to access material in the cup for filling and cleaning (Fig 5c, d). As an added benefit, this means that a single base can be used interchangeably with several 3D printed cups that differ in height and diameter, or that are pre-filled with different fluid samples, as long as they use the same (custom) 3D-printed mating thread. This enables multiple samples to be pre-filled and conditioned with a controlled waiting time and thermal history, if desired, then mounted directly onto the rheometer just prior to testing.

## IV.    RESULTS

Here, we compare measurements of viscous Newtonian oils and a simple yield stress fluid made using our family of vanes to measurements made with standard roughened cone-and-plane fixtures (which we assume to provide true reference values). Although flow curves and viscosity measurements of yield stress fluids often are reported as single values and single curves, the measurements in fact are inexact and repeated measurements commonly vary by more than a few percent for challenging materials. Our aim is to unambiguously compare the performance of a series of vane tools with repeated measurements made using standard cone-and-plate tools to quantify mean error and variability. We use these results also to compare performance among vane designs. Having determined the optimal designs, we then use a printed fractal vane to measure the transient rheological response of multiple TEVP materials.

### A. Measurements of viscous oils

A strain rate sweep experiment was performed on a high viscosity silicone oil ($\mu = 1.0$ Pa.s at 25° C; Fig 6a-c) using a series of vane designs, and compared to the reference measurement made with a cone-and-plate geometry (Fig 6a-c). The measured rotation rate at each torque was converted to shear rate using equation (7) and the torque was converted to stress using equation (12), the *N*-dependent expression.

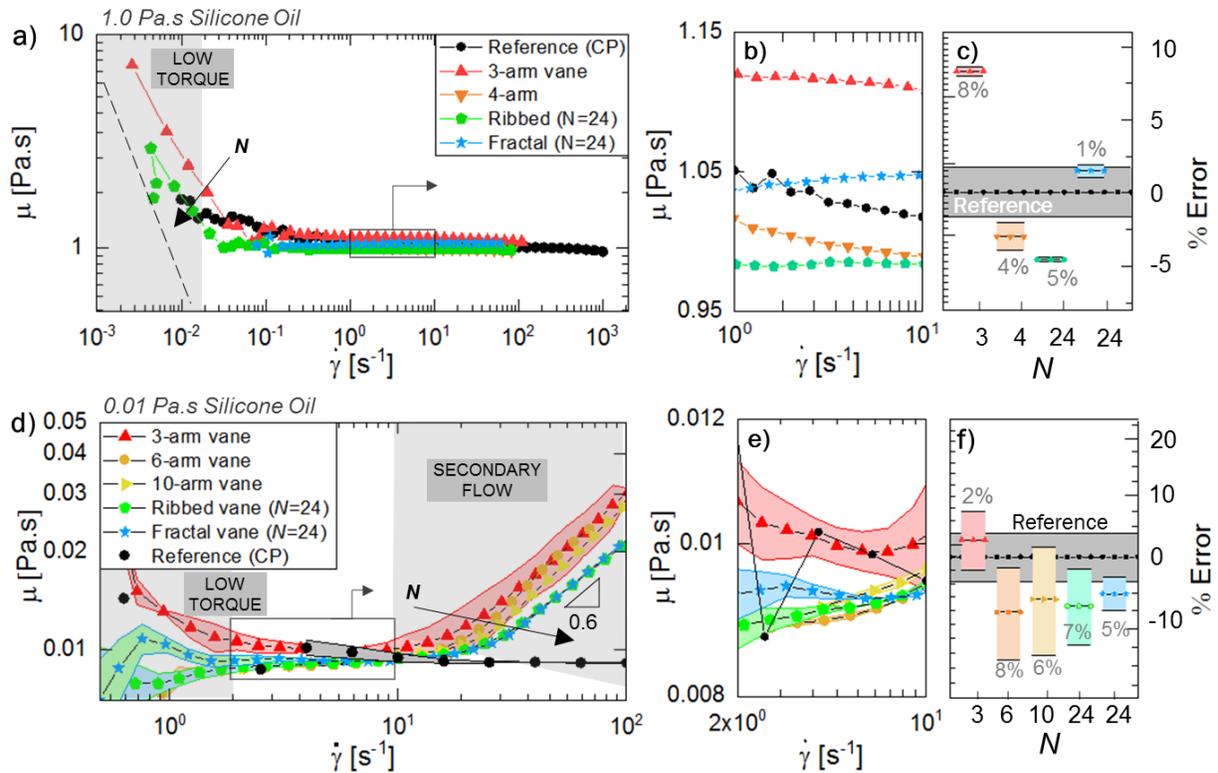

**FIG. 6**: a) Measured viscosity of a viscous Newtonian silicone oil (1.0 Pa.s) measured using several 3D printed vanes (including regular $N = 3$ and $N = 4$ arm vanes plus a 24-armed fractal vane and a ribbed cylinder (see Appendix C for details)), and (b) an expanded view of $1 < \dot{\gamma} < 10 \text{ s}^{-1}$. (c) Viscosity values measured for each geometry compared to the reference measurement made with a cone-and-plate geometry, with the mean percent difference labeled, averaged for three or more measurements for the shear rates in (b). (d) Measured viscosity of a lower viscosity Newtonian silicone oil ($\mu = 0.010$ Pa.s) measured by several 3D printed vanes ($N = 3$, 6, 10, 24-arm fractal, and 24-arm ribbed cylinder). Data is shown with lines indicating mean values and clouds of similar color representing the standard deviation from three or more repeated measurements. At low shear rates $\dot{\gamma} < 2 \text{ s}^{-1}$, machine limitations of torque sensing caused noise. At high shear rates $\dot{\gamma} \gtrsim 20 \text{ s}^{-1}$, secondary flow cause the torque to increase due to inertial effects, with the effect delayed to higher $\dot{\gamma}$ for vanes with more arms. For shear rates of $2 < \dot{\gamma} < 10 \text{ s}^{-1}$, the viscosity was approximately constant for all vanes, and so results are inter-compared in this region. (e) Enlarged view of the region within $2 < \dot{\gamma} < 10 \text{ s}^{-1}$ together with percentage error. (f) Measurement range shown for each geometry compared to the reference measurement made with a cone-and-plate geometry, with the mean percent difference labeled, averaged for three or more measurements over the range of shear rates shown in (e).

When viewed on rheologically-typical log-log axes (such as Fig 6(a)), the results all appear to overlap well, provided the tests are performed above the minimum torque range of the instrument. To investigate the geometric variations we need to look much more closely as shown in Fig 6(b). Here we can see that viscosity measurements using vanes were within 8% of the cone-and-plate reference measurement for shear rates $1 < \dot{\gamma} < 10 \text{ s}^{-1}$, with the fractal design having the smallest error at 1.4% (Fig 6e,f). Because of the large torque values for this viscous fluid and the small variance in the shear stress data for a well-behaved Newtonian oil, the viscosity values measured with the 3-arm vane, 4-arm vane, and 24-arm ribbed vane were statistically distinct from the cone-and-plate measurement (t-test; p<0.05), while only the results obtained using the fractal vane ($N$=24) were not statistically different from the reference curve.

Below the range of shear rates used for this comparison, systematic errors must be recognized. At low shear rates, data becomes noisy when the torque measured by the rheometer reaches the minimum limit of sensitivity ($\approx$10μN.m). Using a vane with more arms increases the window accessible for accurate measurement. The useful range of data was expanded in the low shear rate regime as the number of arms on the vane increased (Fig 6a). A higher number of arms $N$ increases the wetted area of contact between the fluid and the fixture, increasing the conversion factor $S_\sigma$ and leading to a larger torque at low shear rates, thus extending the minimum resolvable stress.

The measurements were repeated using a low viscosity silicone oil (μ = 0.010 Pa.s; Fig 6d-f). As before, the minimum torque sensitivity of the rheometer (10 μN.m) was observed to limit the useful data range at low shear rates. In addition, due to the much lower fluid viscosity raising the Reynolds number characterizing the flow, inertial recirculation between vane arms becomes significant at high shear rates. Within these limiting bounds, $2 < \dot{\gamma} < 10^1 \text{ s}^{-1}$, the measured viscosity was nearly constant (within +/- 10%) for all vanes. Using a vane with a larger number of arms expanded the window accessible for accurate measurement in the low shear rate regime as before, as well as in the high shear rate regime as the number of arms on the vane increased (Fig 6a). At high shear rates, the onset of secondary flow is delayed to higher shear rates, most likely due to the smaller internal characteristic lengths, $l \approx 2\pi R_v / N$, of the lid-driven cavities that exist between neighboring vane arms.

For the low viscosity oil, the variation in the average viscosity measured by all tested vanes was within 8% of the true value, and was less than 5% for the fractal design (Fig 6b,c). However, the measured viscosity was not statistically different from the value measured with a cone-and-plate fixture for any vane (t-test; p>0.05).

## A. Measurement of a simple yield stress fluid

Carbopol is a hydrogel composed of crosslinked poly(acrylic acid) microgel particles that swell in neutral or basic pH to form a physically-jammed microstructure and which, providing the preparation is done carefully, convey a remarkably "simple" yield stress behavior that has been well-studied [4,26,63]. Furthermore, we use a Carbopol-based hair gel (Clear Ice Ultra Hold, Ampro Pro Style) that we found to show simple yield stress behavior. Also, this type of material does not exhibit shear banding, making it useful for calibration [7]. We measured the steady-state flow curve with four 3D printed vanes ($N$ = 3, 4, 12-fractal, 24-fractal, as well as with a solid cylindrical bob) and compared measurements with the reference flow curve obtained using a roughened cone-and-plate geometry. The cone and plate were both roughened by attaching a micropatterned adhesive sandpaper (Trizact A5; 3M), which removes the effect of slip for all measurements. For our measurements, the vanes were inserted into the material until a lower gap equal to $R_v$ was obtained between the bottom of the vane and the lower surface of the cup, and similarly a depth of at least $R_v$ of sample was ensured between the top of the vane and the sample surface.

When using the vane fixtures, the torque measured at each rotation rate increases with the number of outer contact points with the fluid, as expected (Fig 7a). We converted the measured rotation rate to the shear rate in the material at the edge of the vane using equation (7). As shown in Fig 7b, the resulting conversion is rate-dependent rather than a single constant factor, but the curves behave similarly for all vanes. To convert the shear rate precisely, we fit each data set to a functional form $M = a + b\Omega^c$ and inserted that formula into equation (7) for conversion. We found other methods, such as pointwise evaluation of derivatives using centered difference formulae, were substantially less accurate - particularly at low shear rates.

Next, torque was converted to stress using either equation (11) (Fig 7c) or equation (12) (Fig 7d). The solid cylindrical bob showed the onset of slip at low shear rates below $\dot{\gamma}$ = 0.3 s$^{-1}$, as detectable by a systematic deviation to lower measured stresses at a given rate. By contrast, the use of any vane geometry eliminated slip, and each measured curve could be fit to a Herschel-Bulkley model with R$^2$>0.97. When using equation (11), it is clear that the flow curves presented in Fig 7(c) fall systematically below the reference curve, with progressively smaller errors as the number of arms $N$ increased. When using equation (12), most curves collapse onto the reference curve so as to be indistinguishable, with the exception of the cylindrical bob due to slip at low shear rates, and the 3-arm vane, which has substantially larger deviations. The errors (with respect to the reference cone-and-plate data) are shown for each curve as a function of shear rate in Fig 7e and 7g, and the mean error averaged over all shear rates are shown in Fig 7f and 7h, respectively, with labels indicating average percent error.

The total error is much lower when the $N$-dependent stress conversion given by for equation (12) is employed, standard errors are only 0.8-3% and the data fall well within the most likely experimental window expected from the cone-and-plate values (grey shaded region in Fig 7f-h). In comparison, the total error obtained from using equation (11), which assumes an axisymmetric stress field, results in standard errors up to 4-23% in the final flow curves. We further observe that this total error decreases systematically as $N$ increases, to a minimum for the fractal vane with $N = 24$ (Fig 7f), indicating that for the 24-arm vane, the stress field is well-approximated by a cylindrical expression, confirming one of our major design objectives. Our results further support the assertion of Keentok [34] that a 3-armed vane is insufficient for accurate measurement of flow curves in yield stress fluids.

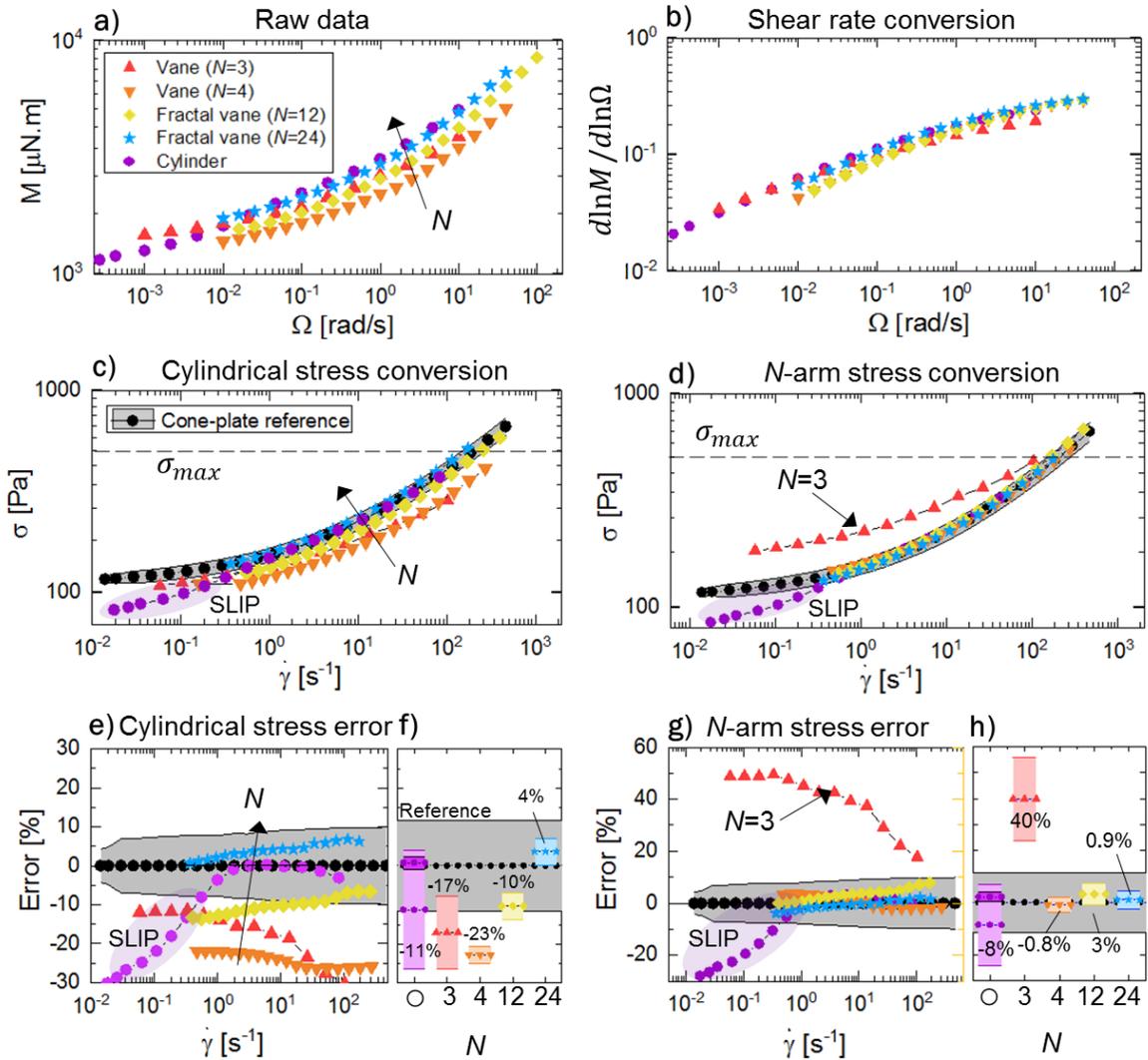

**FIG. 7**: Comparison of vane performance when measuring the flow curve of a Carbopol-based hair gel, a simple (non-thixotropic) yield stress fluid. a) Raw torque vs rotation rate for a range of 3D printed vanes (cylinder, vanes with $N$ = 3, 4, 12, 24 arms), and (b) shear rate conversion factor from equation (7). (c) Final flow curve of shear stress-shear rate after converting stress via equation (11), compared to the cone-plate reference curve (black filled circles). (d) For stress converted via the $N$-dependent equation (12), most data falls on top of the true curve. The exceptions are for the sparse 3-armed vane, and the cylinder which experiences slip at low shear rates. (e,g) The error in the vane data compared to cone and plate reference is shown, along with an error cloud (grey shaded region) showing error from three repeated cone-and-plate measurements of the Carbopol microgel. (f, h) Mean error and the range of error for each vane for three repeated measurements, averaged along the entire curve, where the mean percent error is labeled. The background grey bar shows the range of repeated cone-and-plate reference measurements. The error from the cylinder is represented for both the entire curve including slip (light purple bar) as well as for the partial curve without slip at higher shear rates (dark purple bar). Computing the sample stress using the $N$-dependent expression (h) leads to the lowest overall error for all vanes (excluding the 3-armed vane) with <1% mean error for the fractal and 4-armed vanes, indicating the universal applicability of this conversion expression (equation (12)).

## B. Direct comparison of torque/stress conversion factors with models

In order to determine the best expression to predict the torque/shear stress conversion factors, we calculated the optimal value of $S_\sigma$ to translate the flow curves for a series of fluids exactly onto the reference curve with zero average error over a range of shear rates $10^{-1} < \dot{\gamma} < 10^1 \text{ s}^{-1}$. In Fig 8, we plot these values against the values predicted by equations (11)-(12). Here we combine results for a Newtonian fluid (low viscosity silicone oil shown in Fig 6), a simple yield stress fluid (Carbopol data shown in Fig 7), as well as a more complex thixotropic yield stress fluid (mayonnaise), which is discussed in more detail later. We observe best agreement with equation (12), as well as progressively smaller error bars in the data obtained using vanes with a higher number of arms $N$. the convergence of both expressions at high $N$ allows data taken with a 24-armed vane to compare well with the expected values of $S_\sigma$ for both expressions.

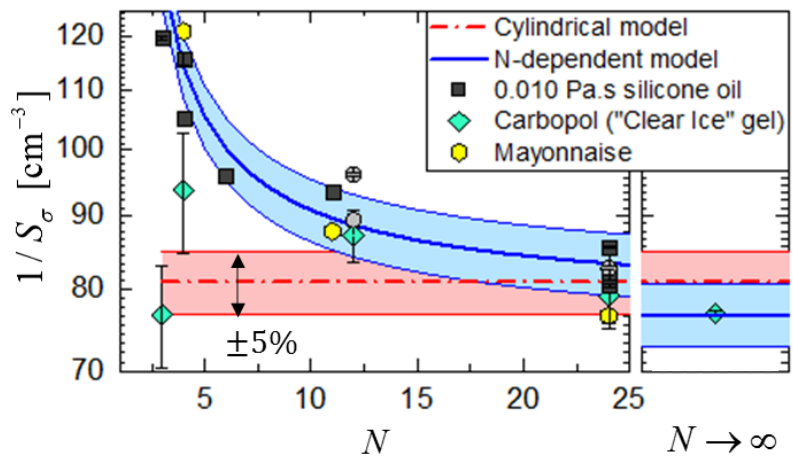

**FIG. 8**. The optimal torque conversion factors calculated from experimental data for three different fluids were compared with the torque conversions factors computed using the two expressions in equations (11) and (12) for a series of $N$-armed vanes. Separately, we also show the conversion factor expected for a Couette "bob" (cylinder) with $N \to \infty$.

## C. Start-up of steady shear flow

Elasto-visco-plastic yield stress fluids also have a characteristic viscoelastic response to start-up of steady shear. Initially, the shear stress increases monotonically with imposed strain before the onset of irreversible plastic yielding and flow. The initial stress growth may be linear, indicating Hookean linear elasticity, or sublinear, indicating a viscoelastic solid response. At a critical apparent strain, the material will begin to fluidize and flow plastically. Eventually the flow will develop a terminal, constant stress at or above the yield stress. If the yield stress fluid is thixotropic, there is often an overshoot before the stress decreases to a constant terminal value [5]. Depending on the imposed shear rate, it will take a different length of time to reach the yielding point. However, in many systems we expect the curves for different shear rates will superpose when plotted against an apparent strain, consistent with a dominant elastic response. Having determined that the $N = 24$ fractal arm vane gives optimal results for a range of fluids, we now perform start-up of steady shear tests on three yield stress materials using a 24-armed fractal vane.

### i. Carbopol

Carbopol was loaded into the textured cup shown in Fig 5 and a constant rotation rate was imposed, ranging from $1.8 \times 10^{-4}$ to $6 \times 10^{-1}$ rad/s, corresponding to shear rates in the range $4 \times 10^{-3}$ to $8\ \mathrm{s}^{-1}$. The rotation rate was imposed for a set time that varied inversely with the rotation rate and increased sequentially in steps, as shown in Fig 9a, with a 10 s waiting time between steps. In response to a single rotation rate, the stress initially grows linearly in time elastically followed by a plastic flow regime as shown in Fig 9b. When this transient stress response is plotted as a function of apparent strain, $\theta = \Omega t$, the curves superpose for all shear rates, as shown in Fig. 9c. The initial deformation is nearly linear in applied strain, allowing us to calculate a shear modulus of $G = 810$ Pa from the average slope of the elastic response, $\Delta \sigma / \Delta \theta$, for all curves. At a critical apparent strain, $\theta_y \approx 0.2$, the material yields and undergoes plastic deformation, reaching a terminal stress that is nearly independent of shear rate over three decades of deformation rate. This is because the very low shear rates applied correspond to the stress plateau region shown in Fig. 7b so $\sigma_w \approx \sigma_y = 110\ Pa$ at long times for all of these deformation rates. In response to increasing the imposed rotation rate $\Omega$, the terminal stress increases slowly and monotonically with rotation rate as expected, but all curves reach the plastic flow regime with a terminal stress within the range $122 \le \sigma^+(t \to \infty) \le 182$ Pa.

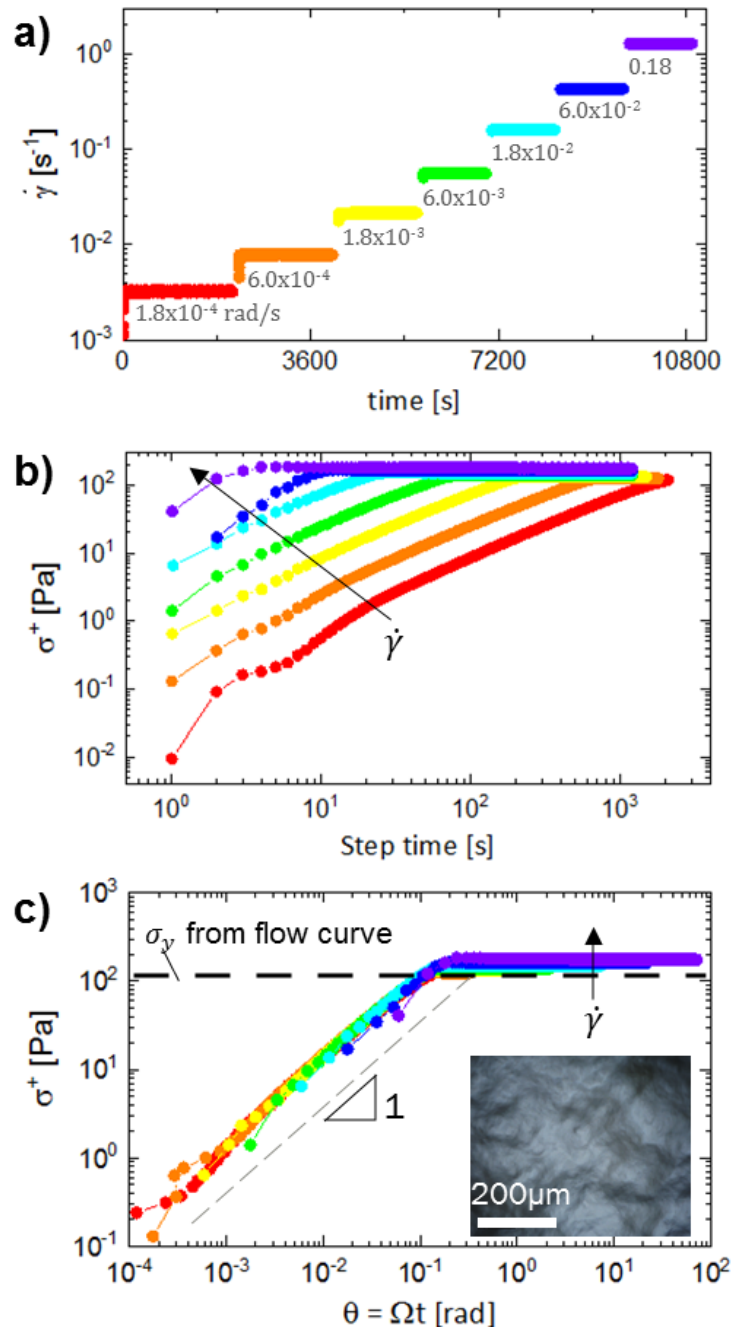

**FIG. 9**. Transient yielding behavior of a Carbopol-based hair gel shows simple yield stress behavior in start-up of steady shear flow tests using a fractal vane ($N = 24$). (a) A series of shear rates were applied, with 1000-2000 s per step, to understand the transient response of the Carbopol. (b) For all imposed rates, the Carbopol shows an initial elastic response followed by plastic yielding at a constant terminal flow stress. (c) When the data are plotted using the apparent strain $\theta = \Omega t$ rather than the experimental step time on the abscissa, they superimpose, revealing a consistent material response. In all cases, the material elastically deforms until it yields at a critical apparent strain $\theta_y \approx 0.2\ rad$ and immediately plateaus at an almost constant value of the stress. The inset shows an optical micrograph of the material microstructure.

## ii. Mayonnaise

Mayonnaise (Hellmann's Squirtable) is a vinegar-and-egg-based emulsion of oil in water (Fig 10a inset). Over time, the fat globules coalesce and particle flocs aggregate, yet imposed stress breaks up structures and/or induces particle migration and mild phase separation [64–66], thus resulting in a thixotropic yield stress response. This makes rheometry of such structured materials very difficult as they show sensitivity to their entire history, including the loading step required to place a sample into the rheometer. For this reason, the mayonnaise samples were held in the original off-the-shelf container for experiments. To investigate the thixotropy of the mayonnaise, a conditioning step was performed to reset the material history after the vane was inserted using a pre-shear rate $\dot{\gamma} = 0.1 \text{ s}^{-1}$ for 10 s, followed by a waiting time, $t_w$, varying from 3 s to 10,000 s. Last, we imposed a constant rotation rate of $1.6 \times 10^{-4}$ rad/s for times $t > t_w$. The resulting material response to this pre-shear history is shown in Fig 10(a). For all waiting times, the initial viscoelastic response is a power law with slope less than unity, indicating strain-dependent plastic losses [67]. Subsequently, the material yields, showing a stress overshoot that depends strongly on the waiting time, $t_w$, before approaching a steady state terminal shear stress. This thixotropic overshoot is indicative of progressive structural buildup following the cessation of preshearing.

The initial power law characteristics of the transient stress growth are common in many microstructured food gels [68]. One way of compactly modeling this response is by using a Scott-Blair fractional element to quantify the rate-dependent material properties. This model has two parameters: a viscoelastic quasi-modulus $\mathbb{G}$, and a fractional exponent $\alpha$ characterizing the order of the fractional derivative [69]. For instance, $\alpha = 0$ indicates a purely elastic response, and $\alpha = 1$ indicates a purely viscous response. The stress-strain relationship for a Scott-Blair element is defined as

$$\sigma_{yx} = \mathbb{G}\frac{d^\alpha \gamma}{dt^\alpha} \tag{14}$$

with the fractional derivative dependent on the exponent $\alpha$. For start-up of steady shear after a waiting time $t_w$, where the apparent strain $\gamma \triangleq \theta = \Omega(t - t_w)$, integration of equation (14) results in

$$\sigma_{yx} = \frac{\mathbb{G}}{\Gamma(1-\alpha)}(t - t_w)^{1-\alpha} \text{ (for } \alpha \neq 0,1) \tag{15}$$

where $\Gamma(\cdot)$ is the gamma function. This time-dependent relationship was fitted to the transient stress growth curves shown in Fig 10(a) for times $t_w \leq t \leq t_{lin}$ where $t_{lin}$ is the point at which the stress first reaches a value equal to its terminal stress. For more details on this fractional model and related analysis, refer to [67–70]. The value of the quasi-property, or scale factor $\mathbb{G}$ in equations (14)-(15) increases from $\mathbb{G} = 600$ Pa.s$^\alpha$ at $t_w = 3$s to $\mathbb{G} = 1080$ Pa.s$^\alpha$, beyond which it plateaus for $t_w \geq 100$ s (Fig 10b). The fractional exponent similarly decreases from 0.46 to 0.14, beyond which it plateaus after 100 s, consistent with the increasingly solid-like nature of the material that occurs during restructuration. Similarly, the yield strain (at which $\sigma = \sigma_{peak}$) increases slightly for all measurements with increasing waiting time, from $\theta_y = 0.05$ rad to $\theta_y = 0.09$ rad. After the initial viscoelastic buildup of stress, the mayonnaise exhibits a second thixotropic behavior with a stress overshoot, in which the stress increases to a maximum value $\sigma_{peak}$ before decreasing to a terminal asymptotic flow stress that is independent of sample age. The terminal stress was 65-70 Pa for all waiting times. The peak stress $\sigma_{peak}$ was constant at 67 Pa for waiting time $t_{wait} < 100s$ and increased afterwards up to 90 Pa. The overshoot stress compared to the terminal stress, $(\sigma_{peak} - \sigma_\infty)/\sigma_\infty$, is plotted in Fig 10c as a function of waiting time. Comparing the results in Fig 10(b) and 10(c), it is clear that there are distinct timescales for restructuring of the linear viscoelastic solid response and for the rise in the nonlinear overshoot stress in this material, which may both be called "thixotropic" timescales. Dividing the material response into multiple timescale processes has recently been suggested by Wei et al [71].

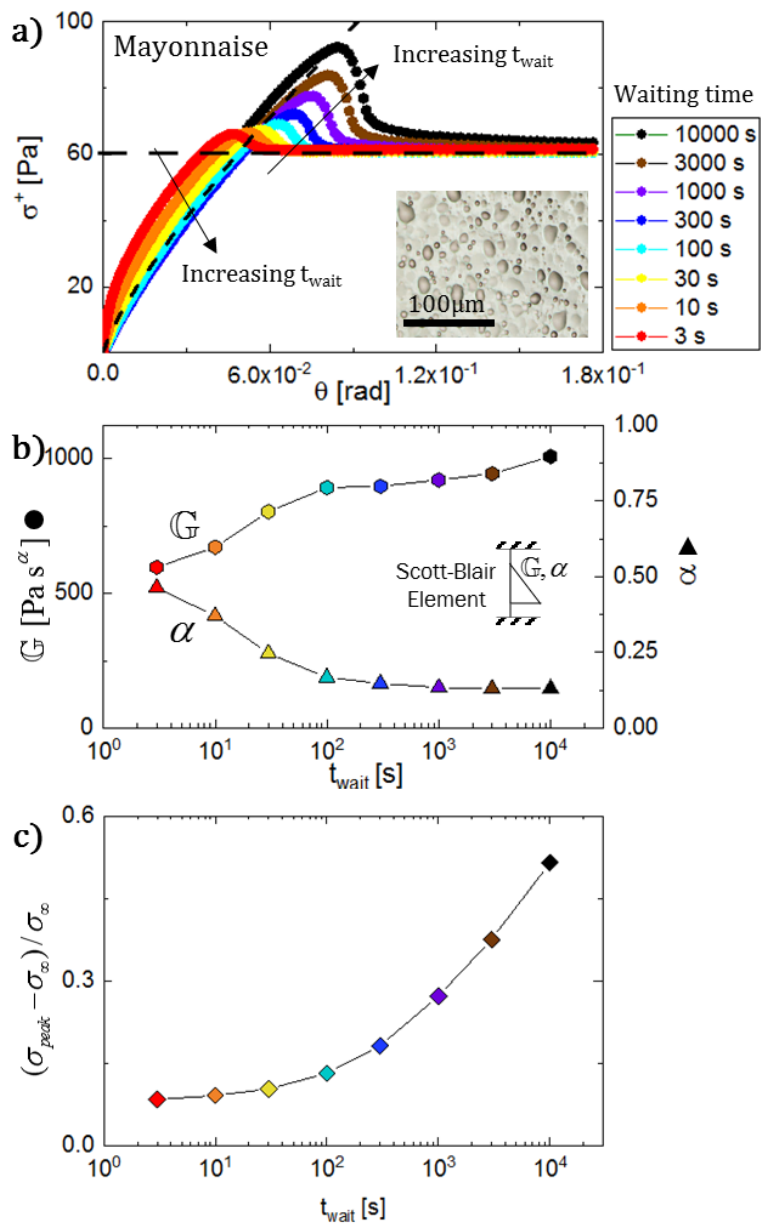

**FIG. 10.** Yielding behavior of mayonnaise (a thixotropic jammed oil-in-water emulsion) measured with a 24-arm fractal vane. A start-up of steady shear test at a constant rotation rate $\Omega = 1.6 \times 10^{-4}$ rad/s was performed after a sequence of pre-conditioning at a shear rate of $0.1\,\text{s}^{-1}$ followed by a waiting time $t_w$. An inset image in (a) shows an optical micrograph of the material microstructure. (b) A fractional model with a single Scott-Blair element was used to describe the viscoelastic material response for times $t_w < t < t_{peak}$. Up until 100 s, the quasi-modulus $\mathbb{G}$ increases and the exponent $\alpha$ decreases rapidly, indicating stronger and more solid-like behavior arising from restructuration. For waiting times $t_w > 100\,\text{s}$, changes in the viscoelastic solid properties were more gradual while (c) the magnitude of the stress overshoot began to increase more rapidly, indicating a complex multiscale reformation of the microstructure between tests.

### iii. Aqueous Battery Slurry

Finally, we measured the yielding characteristics of an aqueous battery slurry composed of Carbopol (Lubrizol, USA; 1 wt%), carbon black (acetylene black, Chevron, USA; 6 wt%), and 7 molar KOH (30 wt%); this mixture results in a paste with pH = 12. The strong alkalinity increases electronic conductivity, improving performance as a battery material [72–74]. It also causes this slurry material to corrode many metals including commercial steel rheometer vanes. While the dispersed carbon black discolored our 3D printed resin, the vane did not show evidence of physical degradation even after 20 hours of experiments.

As with the other materials, a steady rotation rate was imposed ($10^{-4}$ rad/s), without modification between trials. There was no conditioning pre-shear, because no combination of speeds and equilibration times were found to "reset" the material to give a repeatable initial stress response (*i.e.*, the material ages irreversibly). The waiting time was therefore held to a constant value of $t_{wait} = 1000$ s between each test. As shown in Fig 11, the initial stress growth is weakly sublinear in strain but remains almost constant with the number of times the sample is sheared. Fitting the data with equation (15) gives $\mathbb{G} = 430$ Pa.s$^{\alpha}$ and $\alpha = 0.17$. The yield strain, taken at the strain at which stress is a maximum, is $\theta_y \approx 0.1$ rad for the first step, and nearly constant thereafter at $\theta_y \approx 0.06$ rad.

This battery slurry also shows mildly thixotropic behavior with a weak stress overshoot. The stress overshoot remained constant at 20 Pa, while the terminal yield stress decreased with each successive trial from 40 Pa down to 30 Pa, indicating irreversible material aging with repeated shearing. Therefore, this particular slurry recipe would not be stable during use as an electrolyte in a flow battery. Yet, the ability to rapidly and reliably make such measurements using a fractal vane with absence of slip or sample loading artifacts makes the fractal vane useful for assessing development of a functional battery slurry recipe.

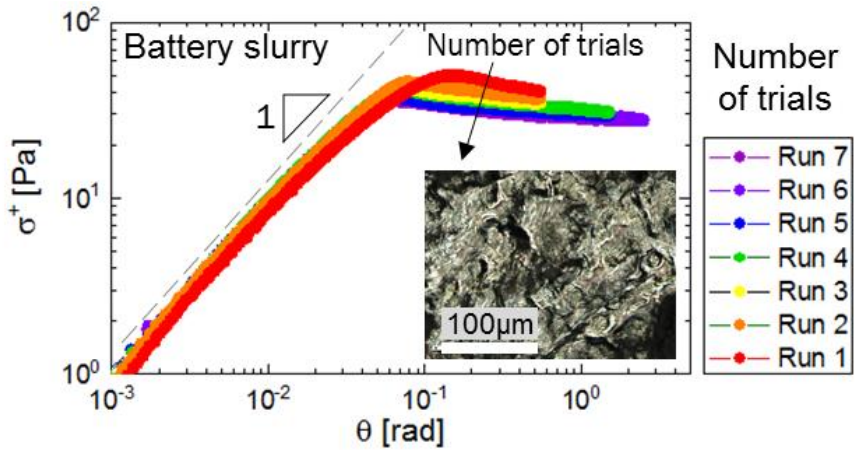

**FIG. 11.** Yielding behavior of a carbon black-based battery slurry measured with a 24-arm fractal vane. The sample permanently degraded with each yielding event, instead of fully recovering, and the stress overshoot decreased with each repeated trial. The inset image shows an optical micrograph of the material microstructure.

## V. CONCLUSION

We have introduced a family of fractal vane geometries, characterized by the number of arms $N$, that can readily be manufactured using cost-effective desktop stereolithographic 3D printing. The design of this new class of fixtures gives a larger surface area-to-volume ratio to the tool, leading to improved axisymmetry of the yield surfaces at the point of yielding and a more accurate determination of the yield stress.

Specifically, we have shown that vanes and textured cups made by desktop stereolithography can be used to obtain rheological measurements that are consistent with "reference" measurements made by machined cone-and-plate geometries with carefully roughened surfaces to eliminate slip. We have also presented expressions for interconverting between measured torque and sample shear stress for a general vane with any number of equally spaced arms, as well as for converting between rotation rate and shear rate for a wide-gapped vane-in-cup configuration. For completeness we repeat these expressions again below.

The torque-to-shear stress conversion is (from equation (12))

$$\sigma = \frac{M}{2\pi R_v{}^2 L \left[ \left(1 - \frac{1.113}{N}\right) + \frac{R_v}{4L}\left(2.75 - \frac{3}{\sqrt{N}}\right) \right]}, \tag{16}$$

and the shear rate conversion at steady state is (from equation (7))

$$\dot{\gamma} = \frac{2\Omega}{d \ln M \ / \ d \ln \Omega} \quad . \tag{17}$$

where $d \ln M \ / \ d \ln \Omega$ is computed from the measured torque-speed curve for the sample under study.

Fabrication of the vanes by 3D printing enabled us to explore various vane designs, and to validate these expressions against experiments. Stress conversion factors derived for an $N$-arm straight-armed vane accurately convert measurements even from a fractal structure with $N$ circumferential points; this is due to the presence of the yield stress, which "cloaks" or hides the internal structure of the vane from the yielded region [75], so that only the number of outer contact edges, and not the internal structure of the vane, affects the yield profile (Fig 2a). The remainder of the fluid plug trapped between the $N$ arms rotates with the vane as a rigid body.

Moreover, vanes made by 3D printing of photopolymer are inexpensive, disposable, and chemically compatible with a wide range of solvents. The total runout of the vanes was measured to be 0.4-0.7 mm at the lower tip of the vane after fitting the printed vanes with an M4 nut as a threaded insert, and without any other post-processing. The printable rheometer coupling is readily adaptable to future design innovations and other rheometer-mounting systems. Such concepts may include designs to minimize moments of inertia, to fit various rheometer and cup geometries, to sharpen or taper the lower surface of each vane arm for easier insertion into soft solids, and to adjust texture, compliance, or porosity for tailoring of specific measurement needs.

We have shown that vanes with consistent, reliable dimensions can be manufactured using a commercially-available desktop 3D printer, yet when using a new 3D printer or a new print material it is always prudent to re-evaluate the chemical compatibility and dimensional accuracy of each printed vane. This can be done using standard alignment calibration fixtures, or by making measurements using a viscous Newtonian calibration oil and comparing these measurements to reference standards, as we have shown in Fig 6.

The average value of the wall shear stress acting on the sample has to be computed from the measured torque, which depends on the (unknown) constitutive model describing the test sample, and so there is unlikely to be a singular "best" vane design for best measurement of all materials. As we and previous researchers have found, viscous Newtonian fluids are more sensitive to internal structural details of the vane which can enhance or inhibit recirculation, whereas yield stress fluids are more sensitive to the number of outer contact edges $N$. As a result, in this work we have considered in detail two key design features; (i) the stress field homogeneity, as shown in Fig 1 and Appendix A, and (ii) the total displaced material as described by the occluded area fraction (OAF), discussed in Appendix C. A schematic representation of the general design tradeoffs we have considered in this work are summarized in Fig 12, together with a qualitative graphical indication (on a sliding scale) of how effective different designs are at addressing each tradeoff.

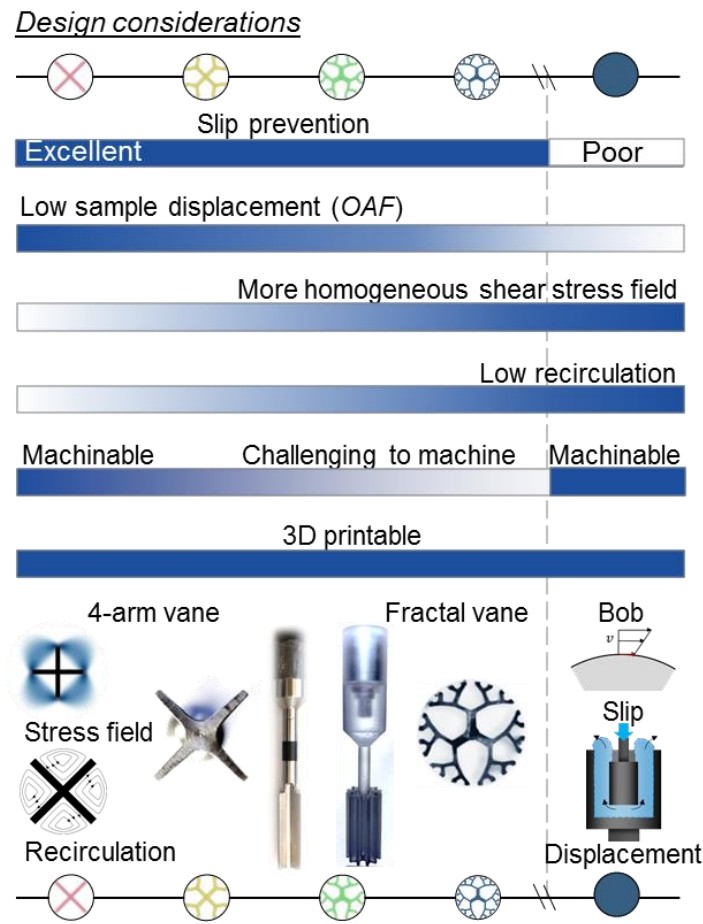

**FIG. 12.** Competing factors affect choice of the rheometric test fixture, *i.e.*, vane, bob, or an intermediate structure. Relative merits are shown with a graded performance scale from excellent (blue) to poor (white). General design criteria include slip mitigation, consideration of sample volume displaced by the vanes, stress homogeneity, recirculation of fluid, and manufacturability. Schematics of these main sources of error for 4-arm vanes and cylindrical bobs are shown above, together with photographs of the 4-arm and 24-arm vanes.

Three-dimensional printing allows us to rapidly explore and screen the performance of a variety of geometric designs. We have found that multipoint fractal designs such as the $N = 12$ and $N = 24$ designs shown in Fig 2 and Fig 12 provide excellent compromises balancing stress homogeneity, low occluded area fraction, and minimal slip artifacts. In addition, our proposed torque-to-stress conversion equation (equation (12)) performs well for a wide range of geometric designs incorporating the role of end-effects and an arbitrary number of arms. This equation, combined with the considerations summarized in Fig 12, allow researchers to identify a bespoke design that addresses specific needs, and tune their own 3D-printed geometries to the specific rheology of the material of interest.

## VI.    ACKNOWLEDGMENTS

C.E.O. was supported by the United States Department of Defense (DoD) through the National Defense Science & Engineering Graduate Fellowship (NDSEG) Program. The authors thank T. M. Narayanan for sample preparation of the battery electrolyte, S. Raayai for help in setting up the 3D printer, and D. Lootens for helpful discussion about calibrating vanes for measuring concrete. Research on the rheometry of yield stress fluids in the Non-Newtonian Fluid Dynamics Group at MIT is supported by a gift from the Procter & Gamble Company.

## VII.    APPENDIX

### A.  Full field stress distribution around *N*-bladed vane

In 1992, C. Atkinson and colleagues analytically calculated the complete two-dimensional stress field around an infinitely long vane with $N$ infinitely thin straight arms drawn from the center to the perimeter of a circle of radius $R = 1$ using the Wiener-Hopf method of solid mechanics [31]. We apply their results here. In particular, for the case of plane strain, the shear stress is

$$\sigma_{r\theta} = \frac{a_{ns}}{\sqrt{\xi}} \left[ \frac{1}{4} \sin\left(\frac{\varphi}{2}\right) \sin(\varphi) \right] \qquad (A1)$$

Here, the coordinate system is a local polar coordinate system with origin at the vane tip and $\varphi = 0$ aligned outwards along the thin straight arm with radius $\xi$ as shown in Fig A1. The prefactor is

$$a_{ns} = \frac{\mu\Omega\sqrt{\alpha/\pi}}{\exp(I/\pi)} \qquad (A2)$$

Here, $\alpha = 2\pi/N$ is the angle between each vane arm, and the integral $I$ is given by

$$I = \int_0^\infty \left( \frac{\log[1 + x\sin(\alpha)\operatorname{cosech}(\alpha x)]}{1 + x^2} \right) dx \qquad (A3)$$

Equation (A2) gives an appropriate viscous scaling for the stress and in Fig A1 we plot contours of $\left| \sigma_{r\theta} / \mu\Omega \right|$ computed from equations (A1-A3).

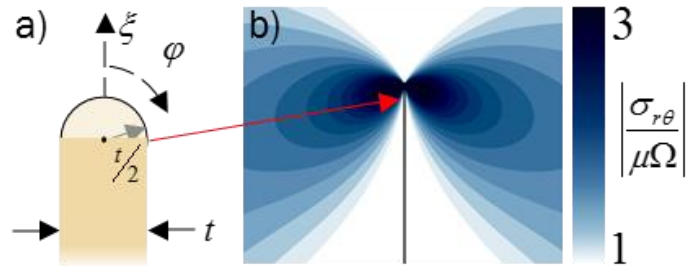

**FIG A1.** Coordinate system of $(\xi, \varphi)$ used here for a vane of thickness $t \ll R_v$. (b) An example of the local dipolar stress field evaluated around a vane tip for a vane with $N = 4$ arms.

In our calculations, we truncated the expression near the vane tip at $\xi = t/2$ where $t$ is the (finite) thickness of an actual vane arm to remove the singularity at $\xi = 0$. In addition, as the stress decays slowly in the far field ($\sigma \propto \xi^{-1/2}$), we truncated the effect of one vane tip at a distance $\xi = 2\pi R_v / N$ to include the effect of each blade to the stress around the first neighboring arm but no further. These truncations generate an integrated shear stress that agrees well with the overall torque equation presented in [31].

## B. Fractal design

The fractal designs considered here are based on finite Bethe lattices. They can be parametrized by the number of initial branches from the center ($Z$) the number of layers or generations ($G$) and the number of new branches emerging from a single branch with each successive generation (which is always two here).

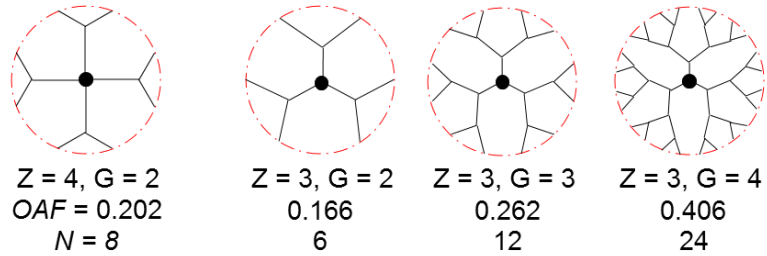

| $Z = 4, G = 2$ | $Z = 3, G = 2$ | $Z = 3, G = 3$ | $Z = 3, G = 4$ |
| *OAF* = 0.202 | 0.166 | 0.262 | 0.406 |
| $N = 8$ | 6 | 12 | 24 |

**FIG B1.** Designs of the fractal vane geometries analyzed here and subsequently 3D printed. (Left) A finite Bethe lattice with $Z = 4$ central branches and $G = 2$ generations. (Right) Lattices with 3 center branches and 2, 3, and 4 generations. The occluded area fraction (OAF) defined in equation (13) are labeled for each design. The values of occluded area fraction are calculated assuming a constant vane thickness of $t = 0.13R_v$.

In addition, the radius ratio of the generations of each subsequent layer can be optimized to minimize the total length of the vane arms. Here, $\beta = r_{i+1} / r_i$ where $r$ as the radius of layer $i$ and $\beta \geq 1$. If $\beta = 1$, we can recover a version of the ribbed cylinder. If $\beta \to \infty$, we can recover the straight-armed vane.

## C. Occluded Area Fraction (OAF) of a vane

For a vane with $N$ straight arms of constant thickness $t$ radiating from the center point, the total occluded area is approximately

$$A = NRt . \tag{C1}$$

Even for thin vane arms with $t << R_v$, if $Nt$ is large, a circular core of the structure is solid to a radius $R_1 = Nt / 2\pi$, and the occluded area is

$$A = \pi R_1^2 + N(R - R_1)t . \tag{C2}$$

Another possible shape is a ribbed cylinder as sketched in Fig 2(d). For this structure, $Z$ branches extend from the central point to $N$ equally spaced points along the outer perimeter connected by chords.. The filled area is minimized by drawing a regular polyhedron with $N$ sides between outer points, connected by one line to the base. The area of that structure is

$$A = Rt\left[2(N-Z)\sin(\pi/N) + Z\right].\tag{C3}$$

For a more general geometry, we employ a Steiner tree algorithm implemented in freely distributed software called GeoSteiner 5.1 [57]. This calculates the global minimum path connecting a series of inputted points by allowing the insertion of additional "Steiner points" along with the input points, which can generate a shorter fully-connected graph than similar problems such as variations on the Traveling Salesman Problem. In this case, the input points are $N$ points evenly spaced around a circle of radius $R = 1$ and one point in the center. The resulting designs are shown in Fig C1.

We can compare and contrast the different designs presented in Fig B1 and Fig C1 by calculating the occluded area fraction (OAF) and plotting this against the number of arms as shown in Fig C2.

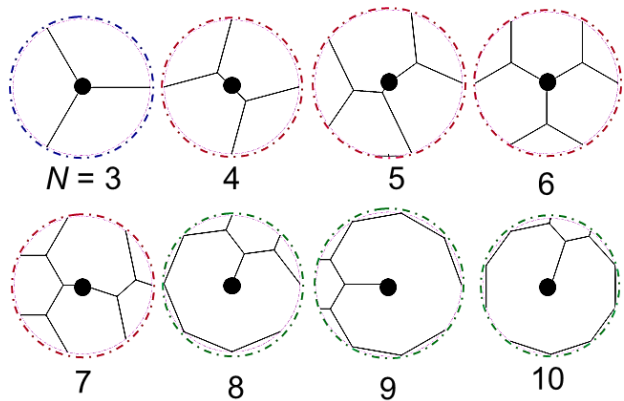

**FIG C1.** Optimal shapes connecting a center point to $N$ equidistant points on the outside of a circle (solved for using a Steiner Tree algorithm [57]).

For low values of $N < \pi + 1$, the OAF of the optimal Steiner tree is most similar to values obtained for vanes with straight radial arms (eg N = 3, shown in Fig C1). For $\pi + 1 < N < 2\pi + 1$, the Steiner tree structure is more fractal-like and has additional branching in the inner spaces of the circle (Fig C1 for $4 \leq N \leq 7$). For $N > 2\pi + 1$, the Steiner tree structure is most consistent with the ribbed cylinder structure ($Z = 1$), with a single connecting support arm extending from the center. Most of the occluded area is located at the perimeter of this design.

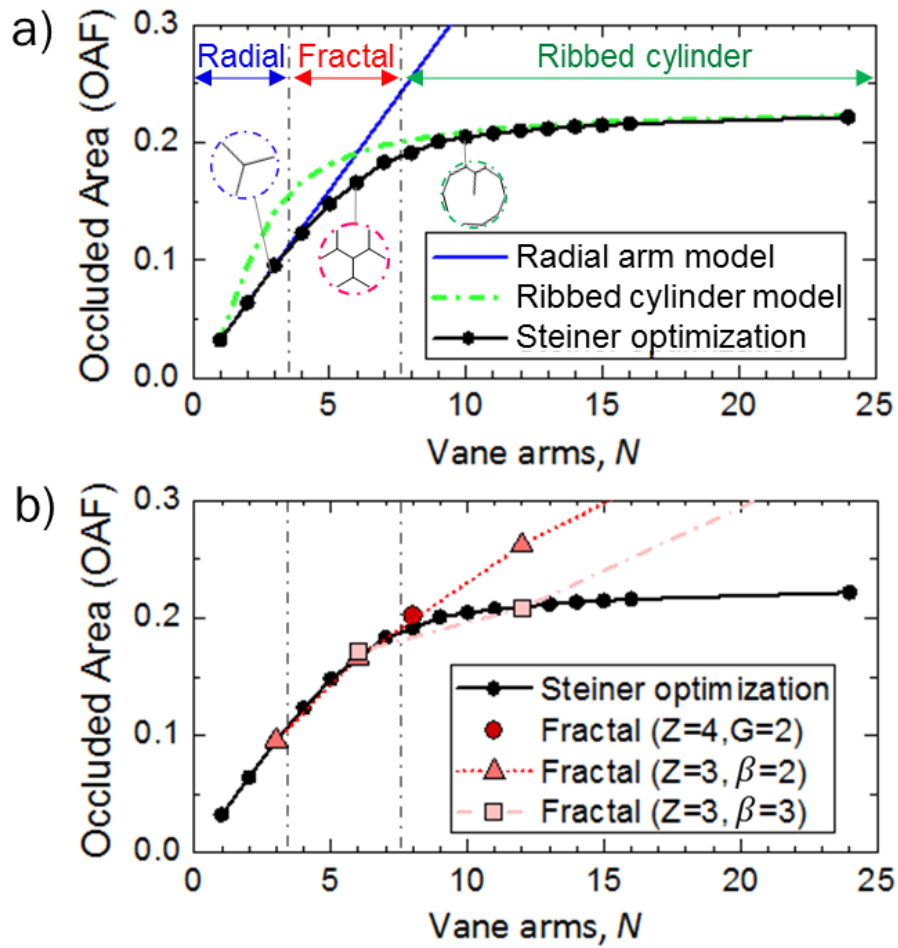

**FIG C2.** (a) Comparison of the Occluded Area Fraction (OAF) for a series of designs of vanes (see equations C2, C3, and (13) in the main text), versus the number of arms, $N$. (b) Comparison of the occluded area for the local optimum shape (as determined by Steiner optimization) as compared to a set of fractal designs such as those shown in Fig B1.

It is also clear from Fig C2(b) that regular stereosymmetric finite Bethe lattice designs can closely approximate the results of Steiner optimization. To further illustrate these choices, we show in Fig C3 four potential designs of an 8-armed vane with corresponding values of $OAF$ listed. While the Steiner tree is optimal, a basic fractal and ribbed structure are only 4% more space-filling, and would be considered "good" designs compared to the straight-armed design, which fills 30% more space. In general, if ease of insertion of the vane is a high priority for material tests, all three styles of design are suitable for $N < 2\pi + 1$, while one would be limited to ribbed cylinder designs and a subset of fractal designs at higher $N$.

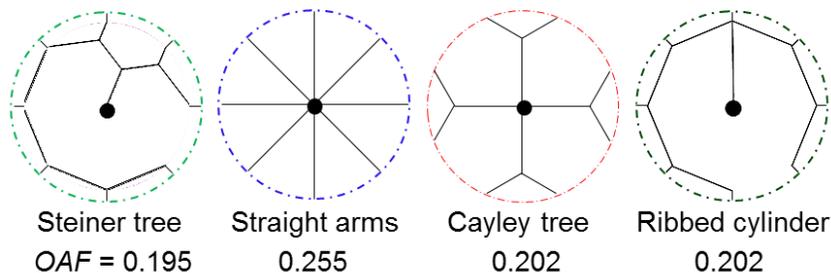

Steiner tree          Straight arms          Cayley tree          Ribbed cylinder
*OAF* = 0.195              0.255                    0.202                    0.202

**FIG C3.** Comparison of four designs of an 8-armed vane using the optimum Steiner tree, straight arms, a fractal structure with 120° angled branches, and a ribbed cylinder. The occluded area fraction (OAF) is calculated for a vane thickness of $t / R_v = 0.13$.

However, the occluded area fraction is not the only essential design criterion. The more structured, space-filling fractals are more suitable than the hollow ribbed structures both for limiting recirculation (by minimizing the distance between any two nearest walls) and for increasing mechanical strength against torsional deformation of the vane features during imposed shearing in stiff materials.

## D. Printing and qualifying a new geometry

The files for the printed vanes are included as supplementary information for this article. They were printed on the Form2 3D printer using Clear v04, Grey v04, White v03 and Black v03 resins from Formlabs, Inc., in an orientation with the vane printing first and the coupling adapter section printing last, without supports (see Fig 3c in the main text). The vanes will have an oblong shape or fail entirely if they are printed at an angle or on their side, due to the nature of stereolithographic printing.

Runout was measured for a series of threaded inserts or other tapping methods to couple the printed geometry to the rheometer, and results of average runout are summarized in Fig D1. The effect of eccentric rotation is an increase in torque measured by the rheometer, which has been calculated previously [76]. For our vanes with a diameter of $2R_v = 15$ mm and thickness $t = 1$ mm, this corresponds to an expected variation in the measured torque that is shown in Fig D2. This is smaller than the typical measurement errors we observed when testing yield stress materials (see Section IV in the main text), even for a runout as large as ~0.5 mm that is visible by eye.

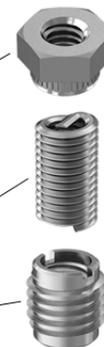

| Coupling type | Runout |
|---|---|
| Commercial TA vane | 0.3 mm |
| **Press-fit nut** | **0.4-0.7** |
| Tapped plastic | 0.9 |
| Press-fit outer hole | 1.0 |
| Helicoil | 1.3 |
| Brass tapping insert | 1.4 |

**FIG D1.** Measured runout for six coupling methods or inserts.

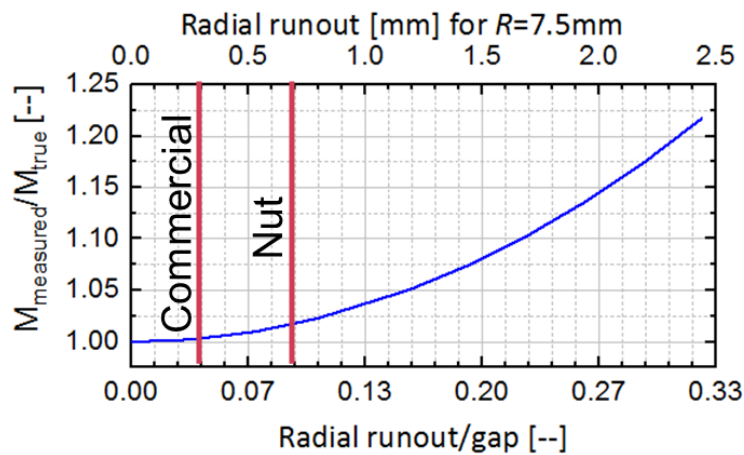

**FIG D2.** Effect of runout on torque measurement, calculated as equation 1C4 in [76]. Lines mark the measured runout for the commercial TA vane and press-fit nut, which give ≲2% error in torque.